\documentclass[twocolumn]{svjour3}          
\smartqed  
\usepackage{graphicx}
\usepackage{amssymb}
\usepackage{mathptmx}      
\usepackage{color}

\newcommand{\mi}{\dot{\mathrm{\i\!\i}}}
\newcommand{\omoon}{{(\!\!|}}

\journalname{CEAS Space Journal}
\begin{document}

\title{HEOSAT: A mean elements orbit propagator program for Highly Elliptical Orbits\thanks{Funded by CNES contract Ref.~DAJ-AR-EO-2015-8181. Preliminary results were presented at the 6th ICATT, Darmstadt, Germany, March 14 -- 17, 2016}
}

\author{Martin Lara         \and
        Juan F.~San-Juan    \and
        Denis Hautesserres
}


\institute{M. Lara \at
              GRUCACI -- University of La Rioja \\
              C/ Madre de Dios 53, Edificio CCT,  \\ 26006 Logro\~no, Spain \\
              Tel.: +34-941-299440\\
              Fax: +34-941-299460\\
              \email{mlara0@gmail.com}           
           \and
           J.F. San-Juan \at
              GRUCACI -- University of La Rioja \\
              \email{juanfelix.sanjuan@unirioja.es}           
           \and
           D. Hautesserres \at
              CCT -- Centre National d'\'Etudes Spatiales \\
	18 Av.~Edouard Belin, \\ 31401 Toulouse CEDEX 4, France \\
              \email{Denis.Hautesserres@cnes.fr}           
}

\date
{CEAS Space Journal (ISSN: 1868-2502, ESSN: 1868-2510) (2018)
\href{https://doi.org/10.1007/s12567-017-0152-x}{doi:10.1007/s12567-017-0152-x} (Pre-print version)}

\maketitle

\begin{abstract}
The algorithms used in the construction of a semi-analytical propagator for the long-term propagation of High\-ly Elliptical Orbits (HEO) are described. The software propagates mean elements and include the main gravitational and non-grav\-i\-ta\-tion\-al effects that may affect common HEO orbits, as, for instance, geostationary transfer orbits or Molniya orbits. Comparisons with numerical integration show that it provides good results even in extreme orbital configurations, as the case of SymbolX.
\end{abstract}

\section{Introduction}
\label{sec:intro}

The subject of analytical or semi-analytical propagation is very old. Since the first analytical orbit propagators based on intermediary solutions to the $J_2$ problem \cite{Sterne1958,Garfinkel1958}, the continuous increase in the accuracy of observations demanded the use of more complex dynamical models to achieve a similar precision in the orbit predictions. In particular higher degrees in the Legendre polynomials expansion of the third-body disturbing function are commonly required (see \cite{Kaufman1981,LaraSanJuanLopezCefola2012}, for instance). Useful analytical theories needed to deal with a growing number of effects, a fact that made that the trigonometric series evaluated by the theory comprised tens of thousands of terms \cite{CoffeyNealSegermanTravisano1995}. 
\par

In an epoch of computational plenty, the vast possibilities offered by special perturbation methods clearly surpass those of general perturbation methods in their traditional application to orbit propagation. Apparently by this reason analytical perturbations have these days been cornered to a downgraded role of providing some insight into the problem under investigation, a task for which a first order averaging is usually considered to be enough, yet the computations of higher orders may provide important details on the dynamics \cite{SanJuanLaraFerrer2006,Lara2008}. However, analytical theories like the popular SGP4 \cite{HootsRoehrich1980} still enjoy a wide number of users mainly involved in catalog propagation duties, a role in which other tools like the Draper Semi-Analytic Satellite Theory \cite{McClain1977,DanielsonNetaEarly1994} or the numeric-analytic theory THEONA \cite{Golikov2012} can compete to numerical integration up to a limited accuracy.\footnote{A partial list of orbit propagators can be found in \href{http://faculty.nps.edu/bneta/papers/list.pdf}{http://fac\-ul\-ty.nps.edu/bneta/papers/list.pdf}, accessed Sep\-tem\-ber 29, 2016.}
\par

On the other hand, new needs in satellite propagation, like the challenges derived of compliance with Space Law, motivate the development of software tools based on analytical or semi-analytical methods, as, for instance, STELA\footnote{\href{https://logiciels.cnes.fr/content/stela?language=en}{https://logiciels.cnes.fr/content/stela}}. In addition, design of end of life disposal strategies may require the long term propagation of thousands of trajectories to find an optimal solution; accurate ephemeris are not needed in the preliminary design and using semi-analytic propagation makes the approach quite feasible \cite{ArmellinSanJuanLara2015}.
\par

Current needs for long-term propagation at the Centre National d'\'Etudes Spatiales motivate the present research. HEOSAT, a semi-analytical orbit propagator to study the long-term evolution of spacecraft in Highly Elliptical Orbits (HEO) is presented. The perturbation model used includes the gravitational effects produced by the more relevant zonal harmonics as well as the main tes\-se\-ral harmonics affecting to the 2:1 resonance of earth's gravitational potential, which has an impact on Molniya-type orbits; the third body perturbations in the mass-point approximation, which only include the Legendre polynomial of second order for the sun and the polynomials from second order to sixth order in the case of the moon; solar radiation pressure, in the cannonball approximation, and atmospheric drag.
\par

The forces of gravitational origin are modeled taking advantage of the Hamiltonian formalism. Besides, the problem is formulated in the extended phase space in order to avoid time-dependence issues. The solar radiation pressure and the atmospheric drag are added as generalized forces. The semi-analytical theory is developed using perturbation techniques based on Lie transforms. Deprit's perturbation algorithm \cite{Deprit1969} is applied up to the second order of the second zonal harmonics, $J_2$. In order to avoid as far as possible the lost of long-period effects from the mean elements Hamiltonian, the theory is corrected by the inclusion of long-period terms of the Kozai-type \cite{Kozai1962,ExertierThesis1988}. The transformation is developed in closed-form of the eccentricity except for tesseral resonances, and the coupling between $J_2$ and the moon's disturbing effects are neglected. 
\par

This paper describes the semi-analytical theory and pres\-ents relevant examples of the numerical validation. An extensive description of the tests performed in the validation of the HEOSAT software is given in \cite{LaraSanJuanHautesserresCNES2016}.
\par

\section{Dynamics of a spacecraft in HEO}

Satellites in earth's orbits are affected by a variety of perturbations of different nature. A full account of this can be found in textbooks on orbital mechanics (see, for instance, \cite{MontenbruckGill2001,Vallado2001}). All known perturbations must be taken into account in orbit determination problems. However, for orbit prediction the accuracy requirements are notably relaxed, and hence some of the disturbing effects may be considered of higher order in the perturbation model. 
\par

Furthermore, for the purpose of long-term predictions it is customary to ignore short-period effects, which occur on time-scales comparable to the orbital period. Thus, in the case of the gravitational potential the focus is on the effect of even-degree zonal harmonics, which are known to cause secular effects. Odd-degree zonal harmonics may also be important because they give rise to long-period effects, whereas the effects of tesseral harmonics in general average out to zero. The latter, however, can have an important impact on resonant orbits, as in the case of geostationary satellites (1 to 1 resonance) or GPS and Molniya orbits (2 to 1 resonance).
\par

The importance of each perturbation acting on an earth's satellite depends on the orbit characteristics, and fundamentally on the altitude of the satellite. Thus, for instance the atmospheric drag, which can have an important impact in the lower orbits, may be taken as a higher order effect for altitudes above, say, 800 km over the earth's surface, and is almost negligible above 2000 km.
\par

\begin{figure}[htb]
\centering
\includegraphics[scale=1.15]{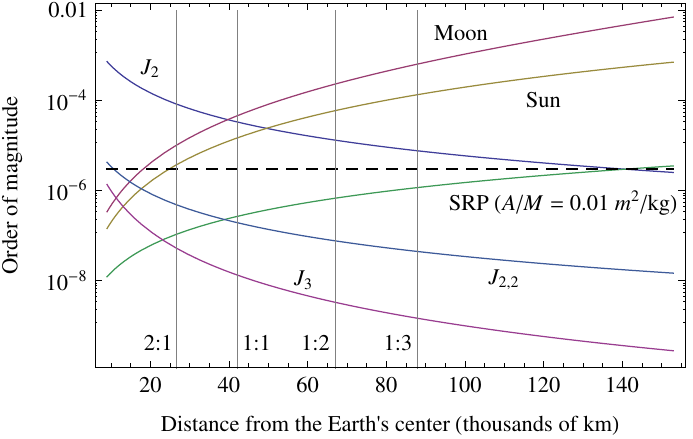} 
\caption{Perturbation order relative to the Keplerian attraction (after \cite{ArmellinSanJuanLara2015}). An area to mass ratio of $0.01\,\mathrm{m^2/kg}$ was taken for estimating the SRP. The horizontal, dashed line corresponds to $3.33\times10^{-6}$ times the Keplerian attraction. }
\label{f:orders}
\end{figure}

A sketch of the order of different perturbations when compared to the Keplerian attraction is presented in Fig.~\ref{f:orders} based on approximate formulas borrowed from \cite[p.~114]{MontenbruckGill2001}. As illustrated in the figure, the non-cen\-tral\-i\-ties of the Geopotential have the most important effect in those parts of the orbit that are below the geosynchronous distance, where the $J_2$ contribution is a first order perturbation and other harmonics cause second order effects. To the contrary, in those parts of the orbit that are farther than the geosynchronous distance the gravitational pull of the moon is the most important perturbation, whereas that of the sun is of second order when compared to the disturbing effect of the moon, and perturbations due to $J_2$ and solar radiation pressure (SRP) are of third order.
\par

Finally, we recall that integration of osculating elements is properly done only in the True of Date system \cite{Efroimsky2005}. For this reason Chapront's solar and lunar ephemeris are used \cite{Chapront-TouzeChapront1988,ChaprontFrancou2003}, which are directly referred to the mean of date in this way including the effect of equinoctial precession.
\par

In the case of a highly elliptical orbit the distance from the satellite to the earth's center of mass varies notably along the orbit, a fact that makes particularly difficult to establish the main perturbation over which to hinge the correct perturbation arrangement. This issue is aggravated by the importance of the gravitational pull of the moon on high altitude orbits, which requires taking higher degrees of the Legendre polynomial expansion of the third-body disturbing function into account. Because this expansion converges slowly, the size of the multivariate Fourier series representing the moon perturbation will soon grow enormously. Besides, for operational reasons, different HEOs may be synchronized with the earth rotation, and hence, being affected by resonance effects. Also, a HEO satellite will spend most of the time in the apogee region, where, depending on the physical characteristics of the satellite, the solar radiation pressure may have an observable effect in the long-term. Finally, the perigee of usual geostationary transfer orbits (GTO) will enter repeatedly the atmosphere, yet only for short periods.
\par

Therefore, a long-term orbit propagator for HEO aiming at describing at least qualitatively the orbit evolution must consider the following perturbations:
\begin{itemize}
\item the effects of the main zonal harmonics of the Geopotential ($J_2$---$J_{10}$ in our case), as well of second order effects of $J_2$;
\item the effects of the tesseral harmonics of the Geopotential that affect the most important resonances, and, in particular, the 2:1 resonance that impacts Molniya orbits
\item lunisolar perturbations (mass-point approximation). The first few terms of the Legendre polynomials expansion of the third-body perturbation are needed in the case of the moon, whereas the effect of the polynomial of the second degree is enough for the sun; 
\item solar radiation pressure effects;
\item and the circularizing effects of atmospheric drag affecting the orbit semi-major axis and eccentricity.
\end{itemize}

In what respects to the ephemeris of the sun and moon, needed for evaluating the corresponding disturbing potentials, the precision provided by simplified analytical formulas is enough for the accuracy requirements of a perturbation theory. For this reason, truncated series from Chapront's solar and lunar ephemeris are taken from \cite{Meeus1998}. The precision of these short truncations is enough for mean elements propagation and notably speed the computations.

\section{Deprit's perturbation approach}

The aim is to find a canonical transformation
\[
({x},{X})\rightarrow({x}',{X}';\epsilon),
\]
where $x\in\mathbb{R}^m$ are coordinates,  $X\in\mathbb{R}^m$ their conjugate momenta, and $\epsilon$ is a small parameter, that converts the Hamiltonian
\begin{equation} \label{Hm0}
\mathcal{H}=\sum_{m\ge0}\frac{\epsilon^m}{m!}\mathcal{H}_{m,0}
\end{equation}
which is written in terms of the ${x}$ and ${X}$, into a new Hamiltonian
\begin{equation} \label{H0m}
\widetilde{\mathcal{H}}=\sum_{m\ge0}\frac{\epsilon^m}{m!}\mathcal{H}_{0,m},
\end{equation}
in the new, prime variables, which is free from short-period terms.
\par

The transformation is derived from a generating function
\begin{equation} \label{Wmp1}
\mathcal{W}=\sum_{m\ge0}\frac{\epsilon^m}{m!}W_{m+1},
\end{equation}
each term of which is computed as a solution of the \emph{homological} equation
\begin{equation} \label{homo}
\mathcal{L}_0(W_m)+\widetilde{\mathcal{H}}_{0,m}=\mathcal{H}_{0,m},
\end{equation}
as follows:
\begin{itemize}
\item terms $\widetilde{\mathcal{H}}_{0,m}$ are known. They are obtained from previous computations based on Deprit's recurrence
\begin{equation} \label{Dr}
\mathcal{H}_{n,q+1}=\mathcal{H}_{n+1,q}+
\sum_{m\ge0}^{n}{n\choose{m}}\,\{\mathcal{H}_{n-m,q};W_{m+1}\},
\end{equation}
where $\{\;;\;\}$ means the Poisson bracket operator.
\item the new term $\mathcal{H}_{0,m}$ is chosen as the part of $\widetilde{\mathcal{H}}_{0,m}$ that is free from short-period terms
\item the term $W_m$ is then solved from Eq.~(\ref{homo}), where
\begin{equation} \label{Lie}
\mathcal{L}_0(W_m)=\{\mathcal{H}_{0,0};W_m\}+\frac{\partial{W}_m}{\partial{t}},
\end{equation}
is customarily known as the \emph{Lie derivative} of $W_m$.
\end{itemize}
\par

After solving the homological equation up to the desired order, the new Hamiltonian in Eq.~(\ref{H0m}) is obtained by changing the old variables in which the terms $\mathcal{H}_{0,m}$ are written by the new ones.
\par

Once the generating function is known, the transformation from old to new variables is itself computed by applying Deprit's recurrence in Eq.~(\ref{Dr}) to the vector functions
\begin{eqnarray*}
{x} &=& \sum_{m\ge0}\frac{\epsilon^m}{m!}{x}_{m,0}({x},{X}), \\
{X} &=& \sum_{m\ge0}\frac{\epsilon^m}{m!}{X}_{m,0}({x},{X}),
\end{eqnarray*}
where ${x}_{0,0}={x}$, ${x}_{m,0}={0}$ ($m>0$), and ${X}_{0,0}={X}$, ${X}_{m,0}={0}$ ($m>0$).
\par

For further details in the procedure the interested reader is referred to Deprit's seminal reference \cite{Deprit1969} or to modern textbooks in perturbation theory or celestial mechanics.

\section{Gravitational perturbations: Hamiltonian approach}

The Hamiltonian Eq.~(\ref{Hm0}) is arranged as follows
\begin{eqnarray*}
\mathcal{H}_{0,0} &=& -\frac{\mu}{2a},
\\
\mathcal{H}_{1,0} &=& -\frac{\mu}{r}\frac{R_\oplus^2}{r^2}C_{2,0}P_2(\sin\varphi) ,
\\
\mathcal{H}_{2,0} &=& 2(\mathcal{Z} + \mathcal{T} + \mathcal{V}_\mathrm\omoon + \mathcal{V}_\odot), \\
\mathcal{H}_{m,0} &=& 0, \qquad m\ge3,
\end{eqnarray*}
where $\mu$ is the earth's gravitational parameter, $(r,\varphi,\lambda)$ are the spherical coordinates of the satellite and $a$ is the semi-major axis of the satellite's orbit, $R_\oplus$ is the earth's equatorial radius, $P_m$ are the usual Legendre polynomials, and $C_{m,0}$ are zonal harmonic coefficients.

\subsection{Geopotential second order perturbations}

The second order Hamiltonian term $\mathcal{H}_{2,0}$ comprises the second order perturbations of gravitational origin. Thus, the higher order terms of the zonal potential are
\begin{eqnarray*}
\mathcal{Z} &=& -\frac{\mu}{r}\sum_{m\ge3}\frac{R_\oplus^m}{r^m}C_{m,0}P_m(\sin\varphi).
\end{eqnarray*}
The tesseral disturbing function is
\begin{eqnarray} \label{tesseral}
\mathcal{T} &=& -\frac{\mu}{r}\!\sum_{m\ge{2}}\frac{R_\oplus^m}{r^m} \\ \nonumber
&& \times\sum_{n=1}^m\left[C_{m,n}\cos{n\lambda}+S_{m,n}\sin{n\lambda}\right]P_{m,n}(\sin\varphi)
\end{eqnarray}
where $P_{m,n}$ are associated Legendre polynomials, and $C_{n,m}$ and $S_{n,m}$, $m\ne0$, are the non-zonal harmonic coefficients. Besides, for orbital applications, instead of using spherical coordinates the satellite's radius vector is expressed in orbital elements.
This can be done by first replacing the spherical coordinates by Cartesian ones
\[
\sin\varphi=\frac{z}{r}, \quad\sin\lambda=\frac{y}{\sqrt{x^2+y^2}}, \quad\cos\lambda=\frac{x}{\sqrt{x^2+y^2}}.
\]
Then, the orbital and inertial frame are related by means of simple rotations, to give
\begin{equation} \label{car2orb}
\left(\begin{array}{c} x \\ y \\ z \end{array}\right)=
R_3(-\Omega)\,R_1(-I)\,R_3(-\theta)\left(\begin{array}{c}r \\ 0 \\ 0 \end{array}\right)
\end{equation}
where $R_1$ and $R_3$ are usual rotation matrices,
\begin{eqnarray*}
\theta &=& f+\omega, \\
r &=& \frac{p}{1+e\cos{f}},
\end{eqnarray*}
$(a,e,I,\Omega,\omega,M)$ are traditional orbital elements and 
\[
p=a(1-e^2),
\]
is the \textit{semi-latus rectum} of the osculating ellipse.
\par

\subsection{Third-body perturbations}

Under the assumption of point masses, the third-body disturbing potential is
\begin{equation} \label{3b}
\mathcal{V}_\star=
-\frac{\mu_\star}{r_\star}\left(\frac{r_\star}{||\mathbf{r}-\mathbf{r}_\star||}-\frac{\mathbf{r}\cdot\mathbf{r}_\star}{r_\star^2}\right)
\end{equation}
where $\mathcal{V}_\star\equiv\mathcal{V}_\omoon$ for the moon, and $\mathcal{V}_\star\equiv\mathcal{V}_\odot$ in the case of the sun. If the disturbing body is far away from the perturbed body, which will always be the case when dealing with perturbed Keplerian motion, Eq.~(\ref{3b}) can be expanded in power series of the ratio $r/r_\star$
\begin{equation} \label{thirdbody}
\mathcal{V}_\star=-\beta\,\frac{n_\star^2a_\star^3}{r_\star}\sum_{m\ge2}\frac{r^m}{r_\star^m}P_m(\cos\psi_\star),
\end{equation}
where
\begin{equation}
\beta=\frac{m_\star}{m_\star+m},
\end{equation}
and
\begin{equation} \label{cosy}
\cos\psi_\star=\frac{xx_\star+yy_\star+zz_\star}{rr_\star}.
\end{equation}
\par

The semi-analytic theory only considers $P_2$ in Eq.~(\ref{thirdbody}) for the sun disturbing potential, whereas $P_2$--$P_6$ are taken for the moon. Both solar and lunar ephemerides are taken from the low precision formulas given by \cite{Meeus1998} (see also \cite{Meeus1997}).
\par

The Lie transforms averaging is only developed up to the second order in the small parameter, so there is no coupling between the different terms of the disturbing function. Hence, the generating function of the averaged Hamiltonian can be split into different terms which are simply added at the end.
\par

\subsection{Time dependency}

Orbits with large semi-major axis will experience high perturbations due to the moon's gravitational attraction, and may be better described in the restricted three-body problem model approximation than as a perturbed Keplerian orbit. However, since our approach is based on perturbed Keplerian motion, we still take the Keplerian term as the zero order Hamiltonian.
\par

Because sun and moon ephemeris are known functions of time, the perturbed problem remains of three degrees of freedom, but the Lie derivative in Eq.~(\ref{Lie}) must take the time dependency into account, viz.
\[
\mathcal{L}_0(\mathcal{W})\equiv\{H_{0,0};\mathcal{W}\}+\frac{\partial{\mathcal{W}}}{\partial{t}}=n\,\frac{\partial{\mathcal{W}}}{\partial{M}}+\frac{\partial{\mathcal{W}}}{\partial{t}},
\]
where $n$ stands for mean motion.
\par

Dealing explicitly with time can be avoided by moving to the extended phase space. Then, assuming that the semi-major axis, eccentricity, and inclination of the third-body orbits, remain constant
\begin{eqnarray*}
\mathcal{L}_0(\mathcal{W}) &=& n\frac{\partial{\mathcal{W}}}{\partial{M}}
+n_\omoon\frac{\partial{\mathcal{W}}}{\partial{M}_\omoon}
+\frac{\partial{\mathcal{W}}}{\partial\omega_\omoon}\frac{\mathrm{d}\omega_\omoon}{\mathrm{d}t}
+\frac{\partial{\mathcal{W}}}{\partial\Omega_\omoon}\frac{\mathrm{d}\Omega_\omoon}{\mathrm{d}t}
 \\ && 
+n_\odot\frac{\partial{\mathcal{W}}}{\partial{M}_\odot}
+\frac{\partial{\mathcal{W}}}{\partial\omega_\odot}\frac{\mathrm{d}\omega_\odot}{\mathrm{d}t}
+\frac{\partial{\mathcal{W}}}{\partial\Omega_\odot}\frac{\mathrm{d}\Omega_\odot}{\mathrm{d}t},
\end{eqnarray*}
which, in view of the period of the lunar perigee motion of 8.85 years (direct motion) and the period of the lunar node motion of 18.6 years (retrograde motion), and the higher periods in the case of the sun relative orbit, can be safely approximated by
\begin{equation}
\label{Liederivative}
\mathcal{L}_0(\mathcal{W})\approx n\,\frac{\partial{\mathcal{W}}}{\partial{M}}
+n_\mathrm\omoon\frac{\partial{\mathcal{W}}}{\partial{M}_\mathrm\omoon}
+n_\odot\frac{\partial{\mathcal{W}}}{\partial{M}_\odot}.
\end{equation}
\par

\subsection{Third body direction}

For the semi-analytic theory, the time dependency will manifest only in the short-period corrections, which are derived from the generating function $\mathcal{W}$. In view of the form of the Lie derivative in Eq.~(\ref{Liederivative}), the directions of both the sun and the moon must be expressed as functions of the sun and moon mean anomaly, respectively. In the case of the sun, we use
\begin{equation} \label{xyzsun}
\left(\begin{array}{c} x_\odot \\ y_\odot \\ z_\odot \end{array}\right)=
R_1(-\varepsilon)\,R_3(-\lambda_\odot)\,R_2(\beta_\odot)\left(\begin{array}{c}r_\odot \\ 0 \\ 0\end{array}\right),
\end{equation}
where $\varepsilon$ is the mean obliquity of the ecliptic, $r_\odot$ is the radius of the sun, and $\beta_\odot$ and $\lambda_\odot$ are the ecliptic latitude and longitude of the sun, respectively. The sun's latitude can be neglected because it never exceeds $1.2$ arc seconds when referred to the ecliptic of the date, whereas the longitude of the sun is expressed in terms of the mean anomaly using standard formulae.
\par

For the moon, because the lunar inclination to the equator is not constant, the orbit is rather referred to the mean equator and equinox of the date by means of four rotations
\[
\left(\begin{array}{c} x_\omoon \\ y_\omoon \\ z_\omoon \end{array}\right)
=R_1(-\varepsilon)\,R_3(-\Omega_\omoon)\,R_1(-J)\,R_3(-\theta_\omoon)\left(\begin{array}{c}r_\omoon \\ 0 \\ 0\end{array}\right),
\]
where $\theta_\omoon$ is the argument of the latitude of the moon (which is also expressed as a function of the moon's mean anomaly by standard formulas), $J\approx5.\!\!^\circ15$ is the inclination of the moon orbit over the ecliptic, which is affected of periodic oscillations whose period slightly shorter than half a year because the retrograde motion of the moon's line of nodes, and $\Omega_\omoon$ is the longitude of the ascending node of the moon orbit with respect to the ecliptic measured from the mean equinox of date.
\par

Note, however, that in the present stage, the semi-an\-a\-lyt\-i\-cal theory only deals with mean elements and corresponding homological equations of the type in Eq.~(\ref{Liederivative}) do not need to be solved at the second order. Future evolutions of the theory will take the short-period corrections due to third-body perturbations into account.
\par

\section{Averaging zonal terms}

The zonal part of the Hamiltonian is first transformed. Because of the actual values of the zonal coefficients of the earth, the old Hamiltonian is arranged in the form
\begin{eqnarray} \label{H00zonal}
\mathcal{H}_{0,0} &=& -\frac{\mu}{2a} \\ \label{H10zonal}
\mathcal{H}_{1,0} &=& -\frac{\mu}{r}\frac{R_\oplus^2}{r^2}C_{2,0}P_2(s\sin\theta) \\ \label{H20zonal}
\mathcal{H}_{2,0} &=& -2\frac{\mu}{r}\sum_{m\ge3}\frac{R_\oplus^m}{r^m}C_{m,0}P_m(s\sin\theta) \qquad
\end{eqnarray}
where we abbreviate $s\equiv\sin{I}$, and all the symbols, viz. $a$, $r$, $I$, and $\theta$, are assumed to be functions of some set of canonical variables. In particular, the averaging is carried out based on the canonical set of Delaunay variables, which is made of the coordinates $\ell=M$, $g=\omega$, $h=\Omega$, and their conjugate momenta
\[
L=\sqrt{\mu{a}}, \qquad G=L\eta, \qquad H=G\cos{I},
\]
respectively, where
\[
\eta=\sqrt{1-e^2},
\]
is customarily known as the ``eccentricity function''. The Delaunay variables are action-angle variables for the Keplerian motion, and fit particularly well in the construction of perturbation theories for perturbed Keplerian motion.
\par

The homological equation of the zonal problem is obtained from Eq.~(\ref{homo}) as
\begin{equation} \label{homoZo}
-n\frac{\partial{W}_m}{\partial\ell}+\widetilde{\mathcal{H}}_{0,m}=\mathcal{H}_{0,m},
\end{equation}
In consequence, the terms of the generating function are computed from quadratures of the form
\begin{equation}\label{quadrature}
W_m=\frac{1}{n}\int(\widetilde{\mathcal{H}}_{0,m}-\mathcal{H}_{0,m})\,\mathrm{d}\ell.
\end{equation}
\par

\subsection{Elimination of the parallax}

It is known that the removal of short-period terms from the Hamiltonian is facilitated to a large extent by a preprocessing of the original Hamiltonian in order to remove parallactic terms \cite{Deprit1981}. That is, by first applying the ``parallactic identity''
\begin{equation} \label{parallactic}
\frac{1}{r^m}=\frac{1}{r^2}\frac{1}{r^{m-2}}=\frac{1}{r^2}\frac{(1+e\cos{f})^{m-2}}{p^{m-2}}, \qquad m>2,
\end{equation}
and next selecting $H_{0,m}$ by removing all the trigonometric terms that explicitly contain the true anomaly in the expansion of $H_{m,0}$ as a Fourier series \cite{LaraSanJuanLopezOchoa2013b,LaraSanJuanLopezOchoa2013c}.
\par

\subsubsection{First order}

At the first step of Deprit's recurrence in Eq.~(\ref{Dr}), the known terms are simply $\widetilde{\mathcal{H}}_{0,1}\equiv\mathcal{H}_{1,0}$, where
\begin{eqnarray*}
\mathcal{H}_{1,0} &=&
C_{2,0}\Big\{(4-6s^2)(1+e\cos{f})+3s^2\Big[e\cos (f+2\omega) \\
&&  +2\cos(2f+2\omega)+e\cos(3f+2\omega)\Big]\Big\}\frac{R_\oplus^2}{r^2}\frac{n^2p^2}{8\eta^6}
\end{eqnarray*}
Then, the new Hamiltonian term $\mathcal{H}_{0,1}$ is selected as,
\begin{equation} \label{H01parallax}
\mathcal{H}_{0,1} =\frac{\mu}{p}C_{2,0}\frac{R_\oplus^2}{p^2}\left(\frac{1}{2}-\frac{3}{4}s^2\right)\frac{p^2}{r^2}
\end{equation}
and the generating function term ${W}_1$ must be computed from Eq.~(\ref{quadrature}) with $m=1$. That is,
\begin{eqnarray} \label{W1parallax}
W_1 &=&\frac{1}{n}\int
C_{2,0}\frac{R_\oplus^2}{r^2}\frac{n^2 p^2}{8\eta^6}\Big[
\left(4-6s^2\right)e\cos{f}  \\ \nonumber
&&  \hspace{2.5cm}+s^2\sum_{j=1}^3jE_{1,j}\cos(jf+2\omega)\Big]\mathrm{d}\ell,
\end{eqnarray}
in which
\begin{equation}
E_{1,1}=3e, \qquad E_{1,2}=3, \qquad E_{1,3}=e.
\end{equation}
\par

Equation (\ref{W1parallax}) is solved in closed form by recalling the differential relation
\begin{equation} \label{dMdf}
\mathrm{d}M=(r/p)^2\eta^3\,\mathrm{d}f,
\end{equation}
based on the preservation of the angular momentum of the Keplerian motion. We get
\begin{eqnarray*}
W_1 &=& k+nR_\oplus^2\frac{C_{2,0}}{8\eta^3}\Big[(4-6s^2)e\sin{f} \\
&& \hspace{2cm} +s^2\sum_{j=1}^3E_{1,j}\sin(jf+2\omega)\Big],
\end{eqnarray*}
where $k$ is an arbitrary function independent of $\ell$.
\par

To avoid the appearance of hidden long-period terms in $W_1$, we choose $k$ in such a way that
\[
\frac{1}{2\pi}\int_0^{2\pi}W_1\,\mathrm{d}M=0.
\]
In this way it is guaranteed that $W_1$ only comprises short-period terms. That is
\begin{eqnarray*}
k &=&-\frac{1}{2\pi}nR_\oplus^2\frac{C_{2,0}}{8\eta^3} \int_0^{2\pi}\Big[(4-6s^2)e\sin{f} \\
&& \hspace{3cm} +s^2\sum_{j=1}^3E_{1,j}\sin(jf+2\omega)\Big]\mathrm{d}M.
\end{eqnarray*}
Using, again, Eq.~(\ref{dMdf}) to compute this quadrature, we get
\[
k=nR_\oplus^2\frac{C_{2,0}}{8\eta^3}\frac{1-\eta}{1+\eta}(1+2\eta)s^2\sin2\omega.
\]
Therefore, calling
\[
E_{1,0}=(1+2\eta)\frac{1-\eta}{1+\eta}=\frac{1+2\eta}{(1+\eta)^2}e^2,
\]
the first term of the generating function of the elimination of the parallax simplification is written
\begin{eqnarray*}
W_1 &=& nR_\oplus^2\frac{C_{2,0}}{8\eta^3}\Big[(4-6s^2)e\sin{f} \\
&& +s^2\sum_{j=0}^3E_{1,j}\sin(jf+2\omega)\Big]
\end{eqnarray*}
which is now free from hidden long-period terms.

\subsubsection{Second order}

Deprit's recurrence now gives 
\begin{eqnarray*}
\mathcal{H}_{0,2} &=& \{\mathcal{H}_{0,1},W_1\}+\mathcal{H}_{1,1}, \\
\mathcal{H}_{1,1} &=& \{\mathcal{H}_{0,0},W_2\}+\{\mathcal{H}_{1,0},W_1\}+\mathcal{H}_{2,0}.
\end{eqnarray*}
Hence, the computable terms of the homological equation (\ref{homoZo}) are
\begin{equation} \label{tildeH02}
\widetilde{\mathcal{H}}_{0,2}=\{\mathcal{H}_{0,1},W_1\}+\{\mathcal{H}_{1,0},W_1\}+\mathcal{H}_{2,0}.
\end{equation}
The computation of the Poisson brackets is straightforward even for the partial derivatives of the true a\-nom\-a\-ly, which cannot be explicitly written in terms of the Delaunay variables. For the reader's convenience, a table of partial derivatives is given in Appendix \ref{ch:spd}, where all of them are expressed as functions of classical orbital elements and related functions rather than in Delaunay variables; this way of proceeding eases computations in the whole process by avoiding the appearance of square roots, on the one hand, and retains the engineering insight provided by the orbital elements, on the other.
\par

After carrying out the required operations, parallactic terms are removed from $\widetilde{\mathcal{H}}_{0,2}$ using Eq.~(\ref{parallactic}). Then, $\widetilde{\mathcal{H}}_{0,2}$ is written in the form of a Poisson series, from which $\mathcal{H}_{0,2}$ is chosen by removing the trigonometric terms of $\widetilde{\mathcal{H}}_{0,2}$ that explicitly depend on the true anomaly, to give:
\begin{eqnarray}  \label{H02parallax}
\mathcal{H}_{0,2} &=& 2\frac{\mu}{p}\frac{p^2}{r^2}\sum_{i=3}^{10}J_i^*
-\frac{\mu}{p}\frac{p^2}{r^2}C_{2,0}^2\frac{R_\oplus^4}{p^4}\left\{ \frac{5}{4}-\frac{21}{8}s^2 \right. \\ \nonumber
&& +\frac{21}{16}s^4+\frac{3}{8}\left(c^2-\frac{5}{8}s^4\right)e^2  -\frac{3}{8}e^2s^2\cos2\omega
\\ \nonumber && 
\left. \times \left[\frac{15}{2}-\frac{35}{4}s^2-\left(4-5 s^2\right)\frac{\eta^2}{(1+\eta)^2}\right] \right\} 
\end{eqnarray}
with $c\equiv\cos{I}$ and
\begin{equation} \label{Ji}
J_i^*=C_{i,0}\frac{R_\oplus^i}{p^i}\sum_{j=0}^{\lfloor{i}/2\rfloor-1}e^{l}Q_{i,l}s^{l}B_{i,l}\,\mi^m\exp(\mi\,l\omega)
\end{equation}
where $l=2j+m$, $\lfloor\;\rfloor$ notes an integer division, $m\equiv{i}\bmod2$, and $\mi=(-1)^{1/2}$. Note that Eq.~(\ref{Ji}) applies also to $i=2$; indeed, it is simple to check that
\begin{equation} \label{H01J2*}
\mathcal{H}_{0,1}=\frac{\mu}{p}\frac{p^2}{r^2}J_2^*.
\end{equation}
The eccentricity polynomials $Q_{i,j}$ in Eq.~(\ref{Ji}) are given in Table \ref{t:EPzonal} of Appendix \ref{ap:tables}, while the inclination polynomials $B_{i,j}$ are given in Tables \ref{t:IPzeven}--\ref{t:IPzodd}.
\par

Finally, the simplified Hamiltonian is obtained by replacing the old variables by the new ones in all the $\mathcal{H}_{0,m}$ terms. That is, the new Hamiltonian is obtained by assuming that all symbols that appear in Eqs.~(\ref{H01parallax}) and (\ref{H02parallax}) are functions of the new (prime) Delaunay variables.

\subsection{Delaunay normalization}

After the preparatory simplification, it is trivial to remove the remaining short-period terms from the simplified Hamiltonian
$\mathcal{H}'=\sum(\epsilon^m/m!)\mathcal{H}'_{m,0}$ where $\mathcal{H}'_{0,0}$ is the same as $\mathcal{H}_{0,0}$, $\mathcal{H}'_{1,0}$ is the same as $\mathcal{H}_{0,1}$ in Eq.~(\ref{H01parallax}), and, $\mathcal{H}'_{2,0}$ is the same as $\mathcal{H}_{0,2}$ in Eq.~(\ref{H02parallax}), but all of them expressed in the new, prime variables.
\par

The new Hamiltonian term $\mathcal{H}'_{0,1}$ is chosen by removing the short-period terms from $\mathcal{H}'_{1,0}$, namely
\begin{equation} \label{H01delaunay}
\mathcal{H}'_{0,1}=\frac{1}{2\pi}\int_0^{2\pi}\mathcal{H}'_{1,0}\mathrm{d}M
=\frac{\mu}{p}C_{2,0}\frac{R_\oplus^2}{p^2}\eta^3\left(\frac{1}{2}-\frac{3}{4}s^2\right)
\end{equation}
which was trivially solved using Eq.~(\ref{dMdf}).
\par

The first term of the new generating function is solved by quadrature from the homological equation, from which
\begin{equation} \label{W1Jdelaunay}
W'_1=\frac{1}{n}\left[-\mathcal{H}'_{0,1}\ell+\int\mathcal{H}'_{1,0}\frac{r^2}{p^2}\eta^3\,\mathrm{d}f\right]=\frac{\phi}{n}\mathcal{H}'_{0,1},
\end{equation}
where
\[
\phi=f-M,
\]
is the equation of the center. Since $\phi$ is made only of short-period terms, there is no need of introducing additional integration constants in the solution of $W'_1$.
\par

Analogously to Eq.~(\ref{tildeH02}), the computable terms of the second order homological equation are
\[
\widetilde{\mathcal{H}}_{0,2}'=\{\mathcal{H}_{0,1}',W_1'\}+\{\mathcal{H}_{1,0}',W_1'\}+\mathcal{H}_{2,0}',
\]
from which expression the new Hamiltonian term $\mathcal{H}_{0,2}'$ is chosen by removing the short-period terms.
\par

After performing the required operations and replacing prime variables by new, double prime variables in all the $\mathcal{H}_{0,m}'$ terms, we get the averaged Hamiltonian $\mathcal{H}''=\mathcal{H}_{0,0}'+\mathcal{H}_{0,1}'+(1/2)\mathcal{H}_{0,2}'$, viz.
\begin{eqnarray} \label{zonalaveraged}
\mathcal{H}'' &=& -\frac{\mu}{2a}+\frac{\mu}{p}\eta^3\sum_{i=2}^{10}J_i^* 
+ \frac{\mu}{p}\eta^3C_{2,0}^2\frac{R_\oplus^4}{p^4}\frac{3}{16}\Bigg\{c^2  \\  \nonumber 
&& \times(1-5c^2) -\left(\frac{1}{3}+s^2-\frac{17}{8}s^4\right)e^2 -\frac{\eta}{2}(1-3c^2)^2  \\  \nonumber
&& \left. -\left[\frac{5}{4}(1-7c^2)-\frac{(1-5c^2)\eta^2}{(1+\eta)^2}\right]e^2s^2\cos2\omega \right\}.
\end{eqnarray}
and the orbit evolution is obtained from Hamilton equations
\begin{eqnarray} \label{Heqc}
\frac{\mathrm{d}(\ell'',g'',h'')}{\mathrm{d}t} &=& \frac{\partial\mathcal{H}''}{\partial(L'',G'',H'')}, \\ \label{Heqm}
\frac{\mathrm{d}(L'',G'',H'')}{\mathrm{d}t} &=& -\frac{\partial\mathcal{H}''}{\partial(\ell'',g'',h'')},
\end{eqnarray}
\par

\section{Third-body averaging}

Now, the perturbation Hamiltonian is arranged as
\[
\mathcal{H}_{0,0}=-\frac{\mu}{2a},\quad \mathcal{H}_{1,0}=0, \quad \mathcal{H}_{2,0} = 2(V_\mathrm\omoon + V_\odot)
\]
where the sun and moon potentials are computed from the Legendre series expansion in Eq.~(\ref{thirdbody}).
\par

Because the maximum power of $\chi$ in $P_m(\chi)$ is $\chi^m$, in view of Eqs.~(\ref{thirdbody}) and (\ref{cosy}), we check that the satellite's radius $r$ appears now in numerators, contrary to the Geopotential case where $r$ always appear in denominators. For that reason, the closed form theory for third-body perturbations is approached using the eccentric anomaly $u$ instead of the true one $f$. Hence,
\[
r=a(1-e\cos{u}),
\]
and the Cartesian coordinates of the satellite are expressed in terms of the eccentric anomaly replacing in Eq.~(\ref{car2orb}) the known relations from the ellipse geometry
\begin{equation} \label{ellipse}
r\sin{f}=a\eta\sin{u}, \qquad r\cos{f}=a(\cos{u} - e).
\end{equation}
\par

In view of $\mathcal{H}_{1,0}\equiv0$, we choose $\mathcal{H}_{0,1}=0$ and hence $W_1=0$. Then, the second order of the homological equation (\ref{homo}) is $\mathcal{L}_0(W_2)+\widetilde{\mathcal{H}}_{0,2}=\mathcal{H}_{0,2}$, where $\widetilde{\mathcal{H}}_{0,2}=\mathcal{H}_{2,0}$ from Deprit's recurrence (\ref{Dr}). The new Hamiltonian term $\mathcal{H}_{0,2}$ is chosen by removing the short-period terms from $\mathcal{H}_{2,0}$. Again, this is done in closed-form by computing the average
\begin{equation} \label{H023b}
\mathcal{H}_{0,2}=\frac{1}{2\pi}\int_0^{2\pi}\mathcal{H}_{2,0}\frac{r}{a}\,\mathrm{d}u,
\end{equation}
where we used the differential relation
\begin{equation} \label{dMdu}
\mathrm{d}M=(1-e\cos{u})\,\mathrm{d}u=(r/a)\,\mathrm{d}u
\end{equation}
which is obtained from Kepler equation.
\par

After computing Eq.~(\ref{H023b}) we obtain the long-term Ham\-il\-to\-ni\-an $\mathcal{H}_{0,0}=-\mu/(2a)$, $\mathcal{H}_{0,1}=0$, and
\begin{equation} \label{H02third}
\mathcal{H}_{0,2}=2(na)^2\beta^*\frac{a_\star^3}{r_\star^3}\frac{n_\star^2}{n^2}\sum_{m\ge2}\frac{a^{m-2}}{r_\star^{m-2}}\,\Gamma_{m},
\end{equation}
with the non-dimensional coefficients
\begin{equation} \label{gammam}
\Gamma_{m}=\sum_{j=0}^{\lfloor m/2\rfloor}A_{m,j}
\sum_{l=-m}^{m}P_{m,j,l}(S_{m,l}^\star\cos\alpha+T_{m,l}^\star\sin\alpha),\;
\end{equation}
where $\alpha=(2j+k)\omega+l\Omega$, and $k=m\,\mathrm{mod}\,2$, cf.~\cite{LaraVilhenaSanchezPrado2015}.
\par

The eccentricity coefficients $A_{m,j}\equiv A_{m,j}(e)$, the inclination ones $P_{m,j,l}\equiv P_{m,j,l}(I)$, and the third-body direction coefficients $T^\star_{m,l}\equiv{T}^\star_{m,l}(u^\star,v^\star,w^\star)$, $S^\star_{m,l}\equiv{S}^\star_{m,l}(u^\star,v^\star,w^\star)$, where
\[
u^\star=\frac{x^\star}{r^\star}, \qquad v^\star=\frac{y^\star}{r^\star}, \qquad w^\star=\frac{z^\star}{r^\star},
\]
are given in Tables \ref{t:EP}, \ref{t:P3B2y3}--\ref{t:P602}, and \ref{t:3BP} of Appendix \ref{ap:tables}, respectively. They are valid for both the moon ($\star\equiv\mathrm\omoon$) and the sun ($\star\equiv\odot$) by using the proper third-body direction vector $(u^\star,v^\star,w^\star)$.
\par

The contribution of lunisolar perturbations to the mean elements equations is by adding to Eqs.~(\ref{Heqc})--(\ref{Heqm}) the terms of Eq.~(\ref{H02third}) derived from corresponding Hamilton equations.

\section{Tesseral resonances}

The tesseral potential is no longer symmetric with respect to the earth's rotation axis. Therefore, longitude dependent terms will explicitly depend on time when referred to the inertial frame. 

To avoid the explicit appearance of time in the Hamiltonian, we move to a rotating frame with the same frequency as the earth's rotation rate $n_\oplus$. The argument of the node in the rotating frame is
\[
h=\Omega-n_\oplus\,t,
\]
and, in order to preserve the symplectic character, we further introduce the Coriolis term $-n_\oplus{H}$ into the Hamiltonian. It is then simple to check that $H=\Theta\cos{I}$ still remains as the conjugate momentum to $h$.

Then, the tesseral Hamiltonian is arranged as a perturbation problem in which
\begin{eqnarray*}
\mathcal{H}_{0,0} &=&-\frac{\mu}{2a}-n_\oplus\Theta\cos{I}, \\
\mathcal{H}_{1,0} &=& 0, \\
\mathcal{H}_{2,0} &=& 2\mathcal{T},
\end{eqnarray*}
where, now, $\mathcal{H}_{0,0}$ is the Keplerian in the rotating frame, and the tesseral potential is given in Eq.~(\ref{tesseral}). Now, the Lie derivative in Eq.~(\ref{Lie}) reads
\[
\{\mathcal{H}_{0,0};W\}=-n\frac{\partial{W}}{\partial\ell}+n_\oplus\frac{\partial{W}}{\partial{h}},
\]
and the solution of the homological equation (\ref{homo}) will introduce denominators of the type $(in-jn_\oplus)$, with $i$ and $j$ integers. Therefore, resonances $n/n_\oplus=j/i$ between the rotation rate of the node in the rotating frame and the mean motion of the satellite introduce the problem of small divisors.
\par

In fact, resonant tesseral terms introduce long-period effects in the semi-major axis that may be not negligible even at the limited precision of a long-term propagation. Therefore, these terms must remain in the long-term Hamiltonian. Furthermore, these terms must be traced directly in the mean anomaly, contrary to true anomaly, to avoid leaving short-period terms in the Hamiltonian, which will destroy the performance of the semi-analytical integration. Hence, trigonometric functions of the true anomaly must be expanded as Fourier series in the mean anomaly whose coefficients are (truncated) power series in the eccentricity.
\par

After the short-period terms have been removed from the tesseral Hamiltonian, we return to the inertial frame by dropping the Coriolis term and replacing $h$ by the right ascension of the ascending node (RAAN), in this way the time explicitly appears into resonant terms of the long-period Hamiltonian.
\par

From Kaula expansions \cite{Kaula1966}, we find that the main terms of the Geopotential that are affected by the 2:1 tesseral resonance are
\begin{eqnarray*}
\mathcal{R}_{2:1} &=& 
-\frac{\mu}{a}\frac{R_\oplus^2}{a^2}\Big\{
F_{2,2,0} G_{2,0,-1} \\ \nonumber
&& \times\left[ C_{2,2}\cos(\alpha+2\omega)+S_{2,2}\sin(\alpha+2\omega)\right] \\
&& +F_{2,2,1} G_{2,1,1} (C_{2,2} \cos\alpha+S_{2,2} \sin\alpha) +F_{2,2,2} \\
&& \times G_{2,2,3}\left[C_{2,2}\cos(\alpha+2\omega)+S_{2,2}\sin(\alpha-2\omega)\right]\!\Big\},
\end{eqnarray*}
in which
\[
\alpha=2(\Omega-n_\oplus{t})+M,
\]
is the (slowly evolving) resonant angle,
\begin{equation} \label{Fi2x2}
\begin{array}{rcl}
F_{2,2,0} &=& \frac{3}{4}(1+c)^2, \\[0.5ex]
F_{2,2,1} &=& \frac{3}{2}s^2, \\[0.5ex]
F_{2,2,2} &=& \frac{3}{4}(1-c)^2
\end{array}
\end{equation}
and, up to $\mathcal{O}(e^{16})$
\[
\begin{array}{ccl}
G_{2,0,-1} &=& -\frac{1}{2}e+\frac{1}{16}e^3-\frac{5}{384}e^5-\frac{143}{18432}e^7-\frac{9097}{1474560}e^9 \\[0.5ex]
&&   -\frac{878959}{176947200}e^{11} -\frac{121671181}{29727129600}e^{13} 
  -\frac{4582504819}{1331775406080}e^{15}
\\[1ex]
G_{2,1,1} &=& \frac{3}{2}e+\frac{27}{16}e^3+\frac{261}{128}e^5+\frac{14309}{6144}e^7 +\frac{423907}{163840}e^9 \\[0.5ex]
&& +\frac{55489483}{19660800}e^{11} +\frac{30116927341}{9909043200}e^{13} 
 +\frac{2398598468863}{739875225600}e^{15}
\\[1ex]
G_{2,2,3} &=& \frac{1}{48}e^3+\frac{11}{768}e^5+\frac{313}{30720}e^7+\frac{3355}{442368}e^9 \\[0.5ex]
&& +\frac{1459489}{247726080}e^{11}+\frac{187662659}{39636172800}e^{13} 
+\frac{33454202329}{8561413324800}e^{15}
\end{array}
\]
Other 2:1-resonant terms can be found in \cite{LaraSanJuanFolcikCefola2011}.
\par

For the 1:1 tesseral resonance, we find
\begin{eqnarray*}
\mathcal{R}_{1:1} &=& -\frac{\mu}{a}\frac{R_\oplus^2}{a^2}\Big\{
F_{2,2,0} G_{2,0,0}\Big[C_{2,2} \cos(2\alpha+2\omega)  \\
&& +S_{2,2}\sin(2\alpha+2\omega)\Big] + F_{2,2,1} G_{2,1,2}\\ \nonumber 
&&  \times\left[C_{2,2} \cos2\alpha +S_{2,2} \sin2\alpha\right] \Big\}
\\ \nonumber
&& -\frac{\mu}{a}\frac{R_\oplus^2}{a^2}\Big\{
F_{2,1,0} G_{2,0,-1}\Big[C_{2,1} \sin (\alpha+2\omega) \\
&& -S_{2,1} \cos (\alpha+2\omega)\Big] +F_{2,1,1}G_{2,1,1} \\
&&  \times\left(C_{2,1} \sin\alpha-S_{2,1}\cos\alpha\right) +F_{2,1,2} G_{2,2,3} \\
&& \times\left[C_{2,1} \sin (\alpha-2\omega)-S_{2,1}\cos (\alpha-2\omega)\right]
\Big\},
\end{eqnarray*}
where, now,
\[
\alpha=\Omega-n_\oplus{t}+M.
\]
\par

In the particular case of the earth, $C_{2,1}=\mathcal{O}(10^{-10})$ and $S_{2,1}=\mathcal{O}(10^{-9})$. Due to the smallness of these values, corresponding terms are commonly neglected from the resonant tesseral potential $\mathcal{R}_{1:1}$. Therefore, the only needed inclination polynomials are $F_{2,2,0}$ and $F_{2,2,1}$, which were already given in Eq.~(\ref{Fi2x2}), whereas the required eccentricity functions, up to $\mathcal{O}(e^{16})$, are
\[
\begin{array}{rcl}
G_{2,0,0} &=& 1-\frac{5}{2}e^2+\frac{13}{16}e^4-\frac{35}{288}e^6-\frac{5}{576}e^8-\frac{49}{3600}e^{10} \\[0.5ex]
&& -\frac{3725}{331776}e^{12}-\frac{7767869}{812851200}e^{14} -\frac{5345003}{650280960}e^{16}
\\[1ex]
G_{2,1,2} &=& \frac{9}{4}e^2+\frac{7}{4}e^4+\frac{141}{64}e^6+\frac{197}{80}e^8+\frac{62401}{23040}e^{10} \\[0.5ex]
&& +\frac{262841}{89600}e^{12}+\frac{9010761}{2867200}e^{14} +\frac{8142135359}{2438553600}e^{16}
\end{array}
\]
\par

Terms of the Hamilton equations derived from the disturbing functions $\mathcal{R}_{2:1}$ or $\mathcal{R}_{1:1}$, will be added to the evolution equations (\ref{Heqc})--(\ref{Heqm}), only for the propagation of those orbits which experience the corresponding resonance.
\par

\section{Generalized forces}

The evolution equations must be completed by adding to the right hand side of Hamilton equations, Eqs.~(\ref{Heqc})--(\ref{Heqm}), the averaged effects of the generalized forces. Because these effects are derived from Gauss equations, we recall that
\begin{eqnarray*}
\frac{\mathrm{d}L}{\mathrm{d}t} &=& \frac{1}{2}na\frac{\mathrm{d}a}{\mathrm{d}t} \\
\frac{\mathrm{d}G}{\mathrm{d}t} &=& \frac{\mathrm{d}L}{\mathrm{d}t}\eta-na^2\frac{e}{\eta}\frac{\mathrm{d}e}{\mathrm{d}t} \\
\frac{\mathrm{d}H}{\mathrm{d}t} &=& \frac{\mathrm{d}G}{\mathrm{d}t}c-na^2\eta{s}\frac{\mathrm{d}I}{\mathrm{d}t}
\end{eqnarray*}
\par

\subsection{SRP}

Under the simplifying assumption that the solar panels remain oriented to the sun, or that the satellite is a sphere (or ``cannonball''), the perturbing acceleration caused by the solar-radiation pressure
\[
\mathbf{\alpha}_\mathrm{srp}=-F_\mathrm{srp}\mathbf{i}_\odot,
\]
is always in the opposite direction of the unit vector of the sun $\mathbf{i}_\odot$. If, besides, it is assumed that \cite{Kozai1961SAO}, 
\begin{itemize}
\item the parallax of the sun is negligible
\item the solar flux is constant along the satellite's orbit
\item there is no re-radiation from the earth's surface 
\end{itemize}
the magnitude of the SRP acceleration is 
\[
F_\mathrm{srp}=(1+\beta)P_\odot\,\frac{a_\odot^2}{r_\odot^2}\frac{A}{m},
\]
where $\beta$ is the index of reflection ($0 < \beta < 1$), $A/m$ is the area-to-mass ratio of the spacecraft, $a_\odot$ is the semi-major axis of the sun's orbit around earth, $r_\odot$ is the radius of the sun's orbit around earth, and the solar radiation pressure constant at one $\mathrm{AU}$ is $P_\odot\approx4.56\times10^{-6}\,\mathrm{N/m^2}$ \cite[p.~77]{MontenbruckGill2001}, 
\par

The components of $\mathbf{i}_\odot$ in the radial, tangent, and normal directions, respectively, are obtained by simple rotations
\[
\mathbf{i}_\odot=R_3(\theta)\,R_1(I)\,R_3(\Omega)\,R_1(-\varepsilon)\,R_3(-\lambda_\odot)\left(
\begin{array}{c} 1 \\ 0 \\ 0 \end{array} \right).
\]
\par

Then, calling $F=-F_\mathrm{srp}/\mu$ Kozai's analytical expressions for perturbations due to SRP \cite{Kozai1961SAO} are easily recovered from the usual Gauss equations. After averaging over the mean anomaly, which is done in closed form based on the differential relation (\ref{dMdu}), we get
\begin{eqnarray} 
\overline{\frac{\mathrm{d}a}{\mathrm{d}t}} &=& 0
\\ \label{esrpa}
\overline{\frac{\mathrm{d}e}{\mathrm{d}t}} &=&
\frac{3}{4}na^2 \Big\{\sin\omega\Big[(\cos\varepsilon-1)\cos(\lambda_\odot+\Omega) \\ \nonumber
&& -(\cos\varepsilon+1)\cos(\lambda_\odot-\Omega )\Big]+\cos\omega\Big[2s\sin\varepsilon \\ \nonumber
&& \times\sin{\lambda_\odot} +c(\cos\varepsilon+1)\sin(\lambda_\odot-\Omega ) \\ \nonumber
&&+c(\cos\varepsilon-1)\sin(\lambda_\odot+\Omega)\Big] \Big\}\eta F
\\  \label{isrpa}
\overline{\frac{\mathrm{d}I}{\mathrm{d}t}} &=& \frac{3}{4}na^2\frac{e}{\eta}F\cos\omega\Big[
s(\cos\varepsilon+1) \sin (\lambda_\odot-\Omega ) \\ \nonumber
&& -2c\sin\varepsilon\sin{\lambda_\odot}+s (\cos\varepsilon-1) \sin (\lambda_\odot+\Omega)\Big] 
\\ \label{hsrpa}
\overline{\frac{\mathrm{d}\Omega}{\mathrm{d}t}} &=& \frac{3}{4}na^2\frac{e}{\eta}\frac{1}{s}F
\sin\omega\Big[s (\cos\varepsilon+1) \sin (\lambda_\odot-\Omega ) \\ \nonumber
&&-2 c \sin\varepsilon\sin{\lambda_\odot}+s (\cos\varepsilon-1) \sin (\lambda_\odot+\Omega)\Big]
\\ \label{wsrpa}
\overline{\frac{\mathrm{d}\omega}{\mathrm{d}t}} &=& -\frac{3}{4}na^2\frac{F}{e \eta }
\Bigg\{\sin\omega\Big[(\cos\varepsilon+1)\sin(\lambda_\odot-\Omega) \qquad \\ \nonumber
&& \times c -2\left(\frac{e^2}{s}-s\right)\sin\varepsilon\sin{\lambda_\odot}+c(\cos\varepsilon-1) \\ \nonumber
&& \times\sin(\lambda_\odot+\Omega)\Big]+\eta^2\cos\omega\Big[(\cos\varepsilon+1) \\ \nonumber
&& \times\cos(\lambda_\odot-\Omega )+(1-\cos\varepsilon)\cos (\lambda_\odot+\Omega )\Big]\Bigg\} 
\\ \label{lsrpa}
\overline{\frac{\mathrm{d}M}{\mathrm{d}t}} &=& n+
\frac{3}{4}na^2 \frac{e^2+1}{e}F\Big\{\sin\omega\Big[c(\cos\varepsilon+1) \\ \nonumber 
&& \times\sin(\lambda_\odot-\Omega ) +c (\cos\varepsilon-1) \sin (\lambda_\odot+\Omega ) \\ \nonumber 
&& +2 s \sin\varepsilon \sin{\lambda_\odot}\Big] +\cos\omega\Big[(\cos\varepsilon+1) \\ \nonumber
&& \times\cos(\lambda_\odot-\Omega)+(1-\cos\varepsilon)\cos(\lambda_\odot+\Omega)\Big]\Big\}
\end{eqnarray}
\par

\subsection{Atmospheric drag: Averaged effects}

Predicting the atmospheric behavior for the accurate evaluation of drag effects seems naive for the long-term scales of interest in this study. However, the atmospheric drag may dominate over all other perturbations in the case of orbits with low perigee heights, even to the extent of forcing the satellite's deorbit.
\par

The magnitude of the drag force depends on the local density of the atmosphere $\rho$ and the cross-sectional area $A$ of the spacecraft in the direction of motion. The drag force per unit of mass $m$ is
\[
\mathbf{\alpha}_\mathrm{drag}=-(1/2)n_\mathrm{d}\mathbf{V},
\] 
where $\mathbf{V}$ is velocity of the spacecraft relative to the atmosphere, of modulus $V$, we abbreviated
\begin{equation} \label{nd}
n_\mathrm{d}=\rho{B}V>0,
\end{equation}
and $B=(A/m){C}_\mathrm{drag}$, is the so-called ballistic coefficient, in which the dimensionless drag coefficient $C_\mathrm{drag}$ ranges from 1.5--3.0 for a typical satellite. Note that $n_\mathrm{d}\equiv n_\mathrm{d}(t)$.
\par

A reasonable approximation of the relative velocity is obtained with the assumption that the atmosphere co-rotates with the earth. Then, from the derivative of a vector in a rotating frame,
\[
\mathbf{V}=\frac{\mathrm{d}\mathbf{r}}{\mathrm{d}t}-\mathbf{\omega}_\oplus\times\mathbf{r}.
\]
We further take $\mathbf{\omega}_\oplus=n_\oplus\mathbf{k}$, in the direction of the earth's rotation axis, and compute its projections in the radial, normal, and bi-normal directions as
\[
\mathbf{\omega}_\oplus=R_3(\theta)\,R_1(I) 
\left( \begin{array}{c} 0 \\ 0 \\ n_\oplus \end{array} \right).
\]
\par

Then, the velocity components in the radial, normal, and bi-normal direction relative to a rotating atmosphere are
\[
\mathbf{V}=\left(\begin{array}{c} R \\  (\Theta/r)-rn_\oplus\cos{I} \\ rn_\oplus\cos\theta\sin{I} \end{array}\right),
\]
where 
\begin{eqnarray*}
R &=& \frac{\mathrm{d}r}{\mathrm{d}t}=\frac{\Theta}{p}e\sin{f}, \\
\Theta &=& r^2\frac{\mathrm{d}\theta}{\mathrm{d}t}=\sqrt{\mu{p}}. 
\end{eqnarray*}
\par

Models giving the atmospheric density are usually complex. Furthermore, since the atmospheric density depends on the solar flux which is not easily predictable, reliable predictions of the disturbing effects caused by the atmospheric drag are not expected for long-term propagation. Hence, the aim is rather to show the effect that the atmospheric drag might have in the orbit, as opposite from a drag-free model. Therefore, to speed evaluation of the semi-analytical propagator, we take advantage of the simplicity of the Harris-Priester atmospheric density model \cite{HarrisandPriester1962}, which is implemented with the modifications of \cite{LongCappellariVelezFuchs1989}.
\par

After replacing $\mathbf\alpha_\mathrm{drag}$ into Gauss planetary equations, the long-term effects are computed by averaging the equations over the mean anomaly, viz.
\begin{eqnarray} \label{draga}
\overline{\frac{\mathrm{d}a}{\mathrm{d}t}} &=&-\frac{a}{\eta^2}\frac{1}{2\pi}\int_0^{2\pi}n_\mathrm{d} \\ \nonumber
&& \times\left(1+2e\cos{f}+e^2-\frac{n_\oplus}{n}\eta^3c\right)\mathrm{d}M
\\
\overline{\frac{\mathrm{d}e}{\mathrm{d}t}} &=&-\frac{1}{2\pi}\int_0^{2\pi}n_\mathrm{d} \\ \nonumber
&& \times\left[e+\cos{f}-\delta{c}\left(e+\cos{f}-\frac{e}{2}\sin^2f\right)\right]\mathrm{d}M 
\\
\overline{\frac{\mathrm{d}I}{\mathrm{d}t}} &=& -\frac{1}{2}s\frac{1}{2\pi}\int_0^{2\pi}n_\mathrm{d}\,\delta\cos^2\theta\,\mathrm{d}M
\\
\overline{\frac{\mathrm{d}\Omega}{\mathrm{d}t}} &=& -\frac{1}{2}\frac{1}{2\pi}\int_0^{2\pi}n_\mathrm{d}\,\delta\sin\theta\cos\theta\,\mathrm{d}M
\\
\overline{\frac{\mathrm{d}\omega}{\mathrm{d}t}} &=& -c\,\overline{\frac{\mathrm{d}\Omega}{\mathrm{d}t}}
 \\ \nonumber
&& -\frac{1}{2\pi}\int_0^{2\pi}
\frac{n_\mathrm{d}}{e}\sin{f}\left[1-\delta{c}\left(1+\frac{e}{2}\cos{f}\right)\right]\mathrm{d}M 
\\ \label{dragM}
\overline{\frac{\mathrm{d}M}{\mathrm{d}t}} &=& n+\frac{1}{2\pi}\int_0^{2\pi}n_\mathrm{d}\,\frac{e}{\eta}\frac{r}{a}\sin{f}\,\mathrm{d}M \\ \nonumber
&& +\frac{1}{2\pi}\int_0^{2\pi}n_\mathrm{d}
\frac{\eta}{e}\sin{f}\left[1-\delta{c}\left(1+\frac{e}{2}\cos{f}\right)\right]\mathrm{d}M
\end{eqnarray}
in which $\delta=(n_\oplus/n)(r/p)^2\eta^3$.
\par

Both the relative velocity with respect to the rotating atmosphere $V$, and the atmospheric density $\rho$ are naturally expressed as a function of the true anomaly \cite{LongCappellariVelezFuchs1989}, then it happens that $n_\mathrm{d}\equiv n_\mathrm{d}(f)$ from the definition of $n_\mathrm{d}$ in Eq.~(\ref{nd}). Hence, the quadratures above are conveniently integrated in $f$ rather than in $M$ using the differential relation in Eq.~(\ref{dMdf}). Besides, due to the complex representation of the at\-mos\-pher\-ic density, these quadratures are evaluated numerically.
\par

\section{Sample tests}

To illustrate the performance of the mean elements theory we describe two test cases: one for a Molniya-type orbit, and the other for a SymbolX-type orbit, in which the propagations are extended to 100 years.
\par

A full account of the different tests that have been carried out in the development of the HEOSAT software can be consulted in \cite{LaraSanJuanHautesserresCNES2016}. The test cases include GTO, super GTO, and SSTO orbits, as well as Tundra orbits, and orbits of the telescope satellites' missions Integral and XMM-Newton. The semi-analytical theory generally runs one or two orders of magnitude faster than the Cowell integration, although these ratios notably reduce when the atmospheric drag has a non-negligible effect. In spite of that, in all the tested cases HEOSAT runs more than 5 times faster than the numerical integration.
\par

\subsection{Molniya orbit}

The first test presented is for a Molniya type orbit. The initial conditions used in the test correspond to the osculating elements
\begin{equation} \label{Molniya}
\begin{array}{rcl}
a &=& 26554.0\,\mathrm{km} \\
e &=& 0.72 \\
I &=& 63.4 \deg \\
\Omega &=& 0.1 \deg \\
\omega &=& 280 \deg \\
M &=& 0
\end{array}
\end{equation}
The numerical reference has been computed with a Cowell method of order 8, and the integration step size was 30 seconds. In this example, the satellite reaches about the 10\% of the earth-moon distance at apogee, thus suffering moderate third-body perturbations.
\par

\begin{figure}[t]
\includegraphics[scale=0.6]{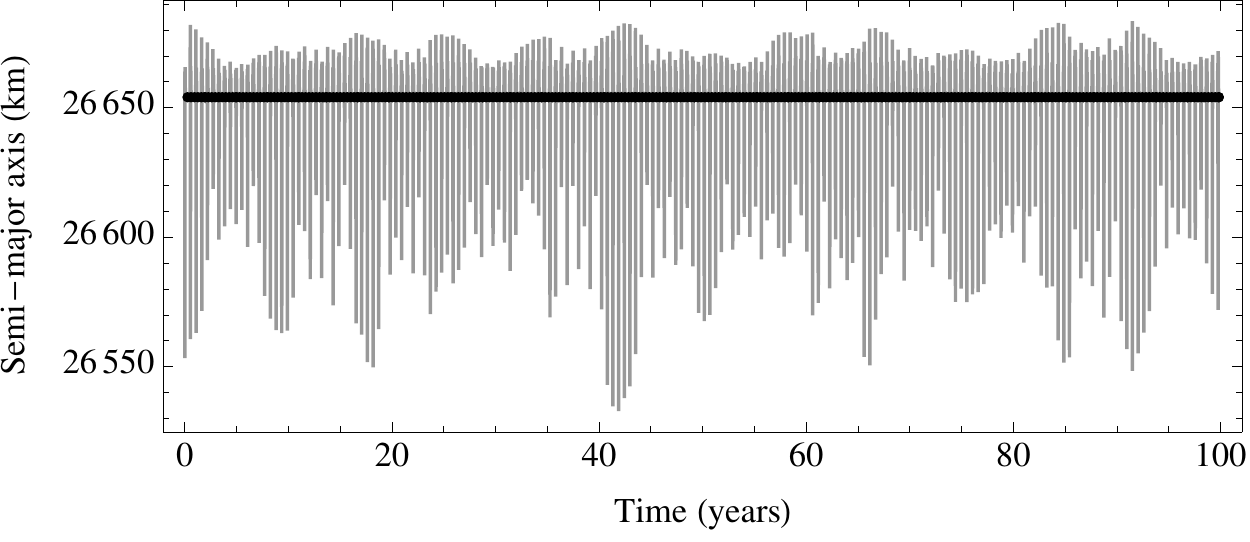} \\
\includegraphics[scale=0.6]{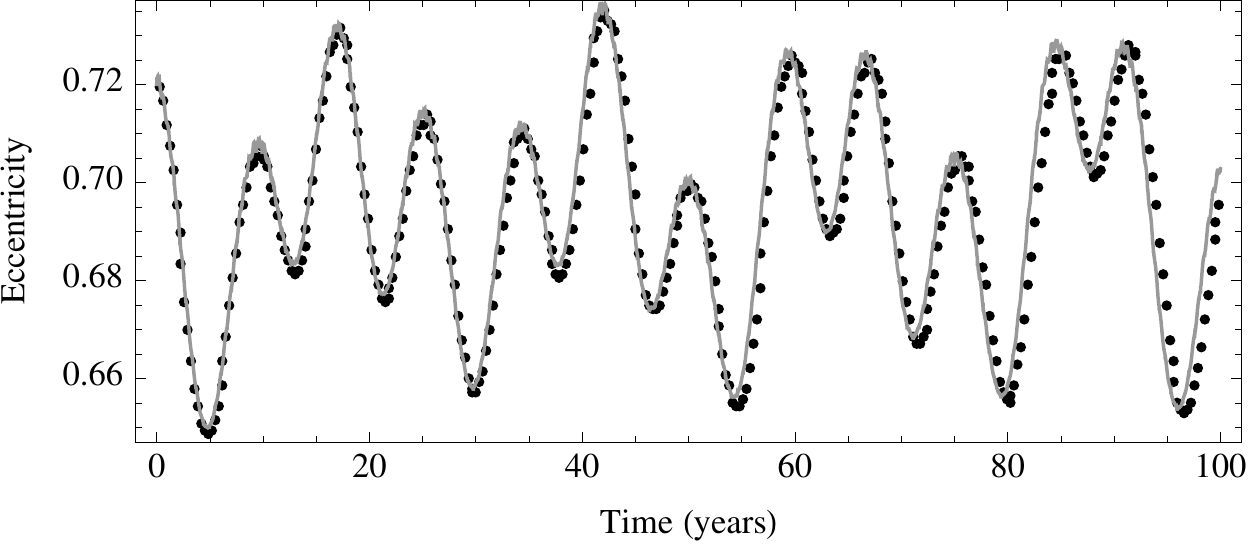} \\
\includegraphics[scale=0.6]{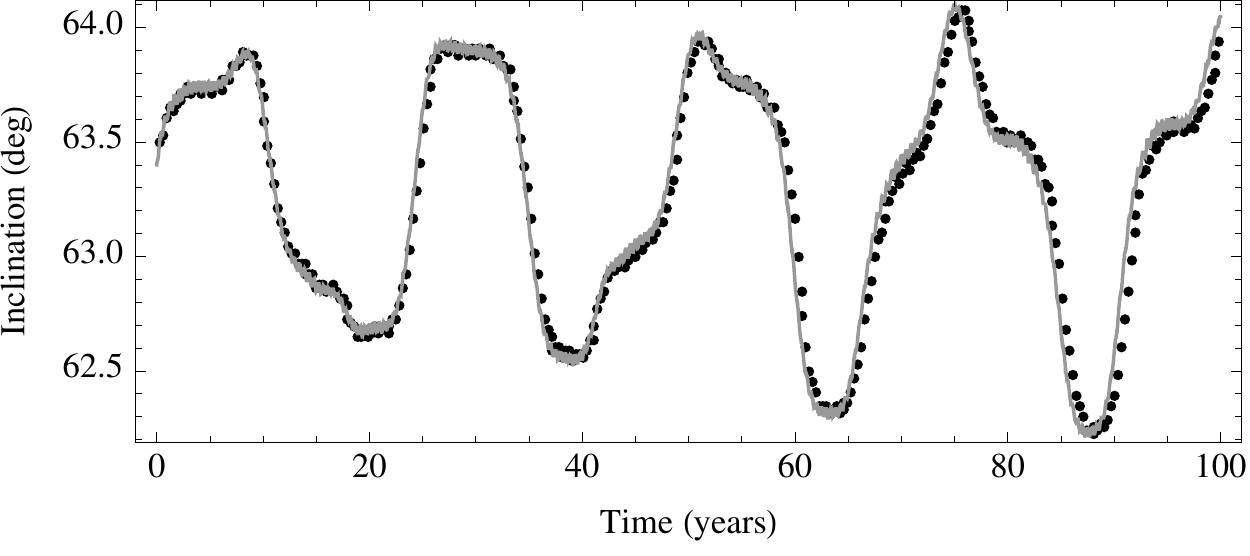} \\
\includegraphics[scale=0.6]{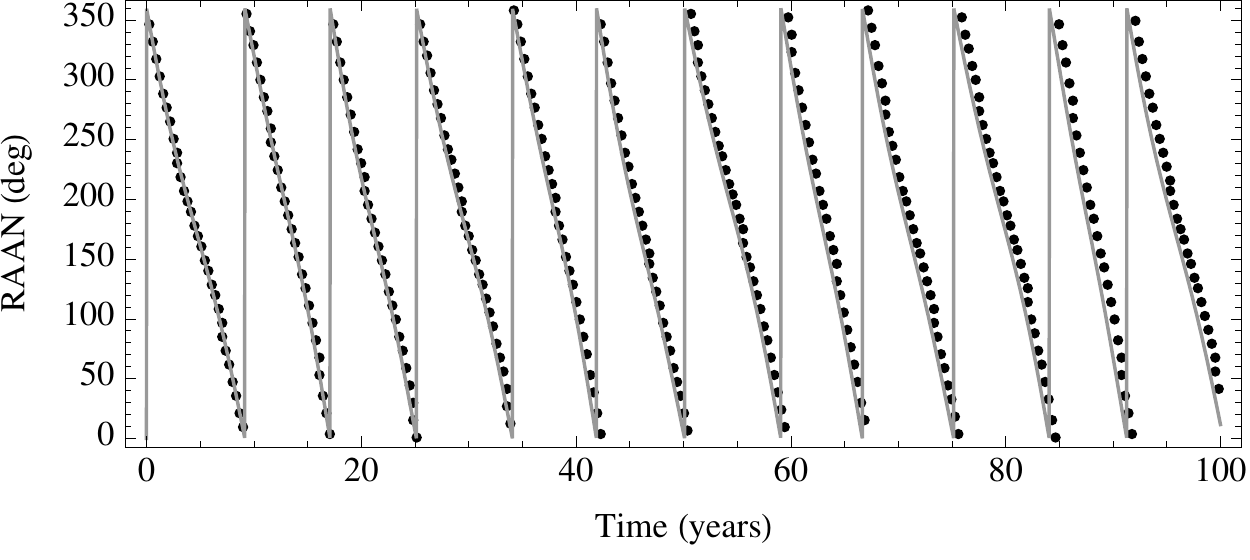} \\
\includegraphics[scale=0.6]{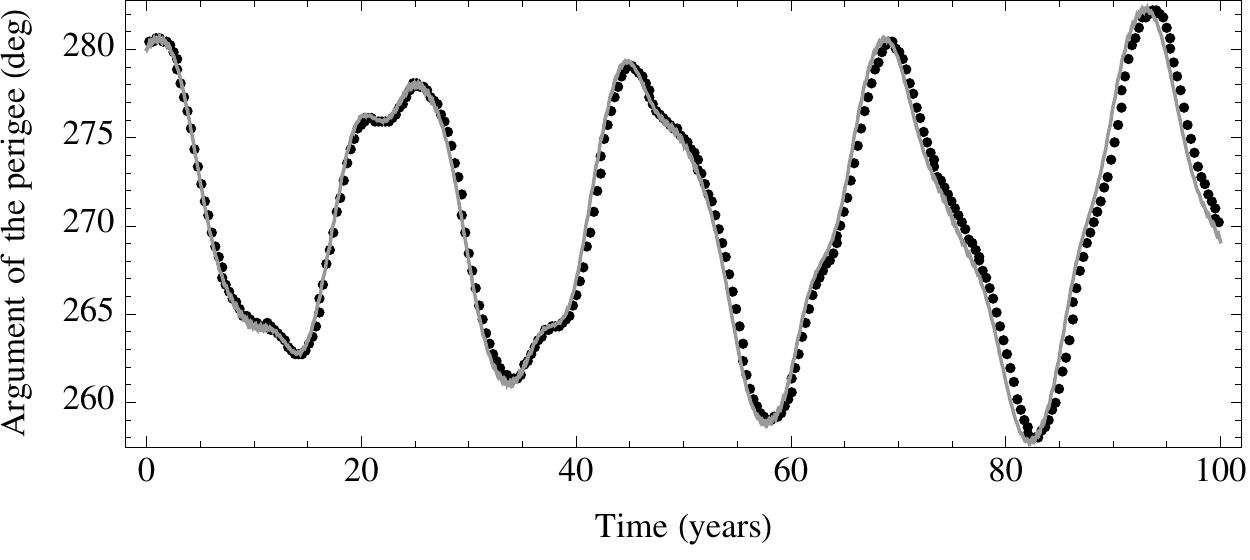} 
\caption{Time history of the orbit elements of the Molniya orbit. Dots: mean elements propagation; gray line: numerical reference}
\label{f:Molniyath}
\end{figure}

The time history of the orbital elements corresponding to these initial conditions is presented in Fig.~\ref{f:Molniyath}. As shown in Fig.~\ref{f:Molniyath}, the numerical reference and the mean elements propagation fit quite well, with a slight shift to the right of the mean elements orbit. This shift is due to the difference between the osculating and mean elements used in both kind of propagations. Indeed, as far as the conversion from osculating to mean elements is not implemented in the current version HEOSAT, the initial mean elements used in launching the semi-analytical propagation do not correspond exactly to the initial conditions used in the propagation of the osculating reference orbit, even though the semi-major axis of the mean elements propagation has been manually adjusted to the mean value of $26653.5$ km, to which the osculating semi-major axis approximately averages.
\par

Better agreement in the comparisons between the mean elements propagation and the numeric reference is expected when the HEOSAT software be completed with the analytic transformation from osculating to mean elements. Second order short-period corrections of the initial semi-major axis value related to this transformation are known to have non-negligible effects in the computation of the mean semi-major axis, and are of similar importance to the first order periodic corrections of the osculating to mean conversion of the other elements \cite{HautesserresLaraCeMDA2016,BreakwellVagners1970}.
\par

\begin{figure}[t]
\includegraphics[scale=0.6]{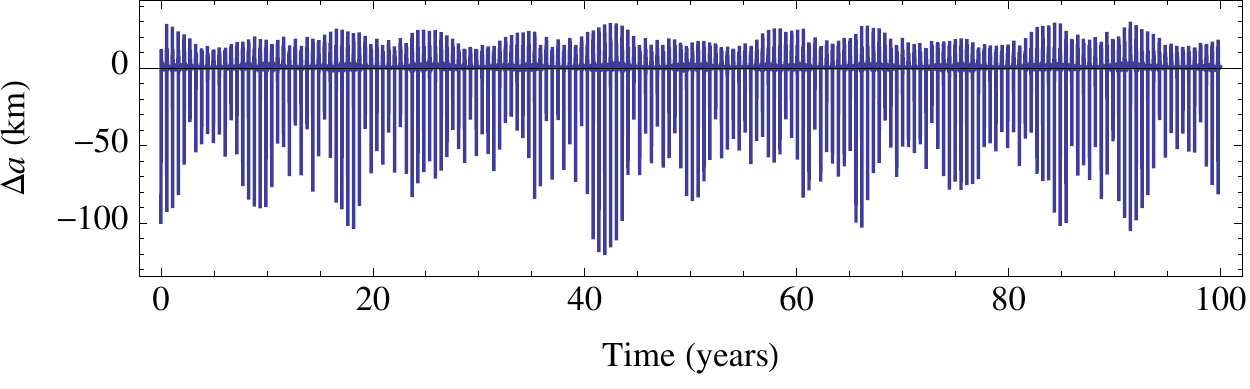} \\
\includegraphics[scale=0.6]{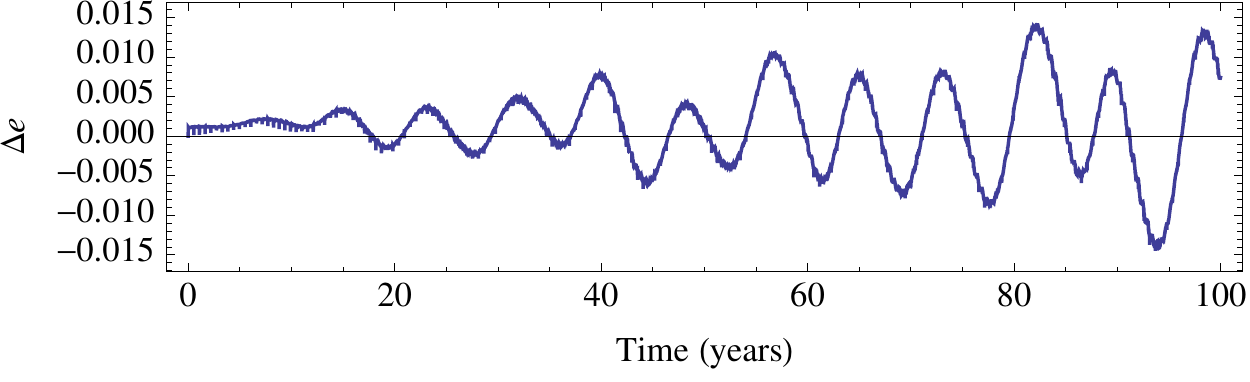} \\
\includegraphics[scale=0.6]{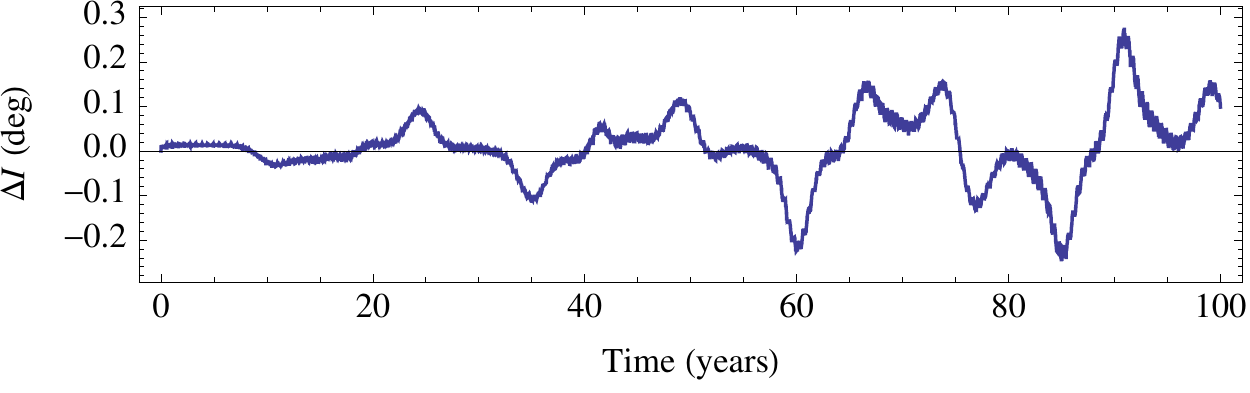} \\
\includegraphics[scale=0.6]{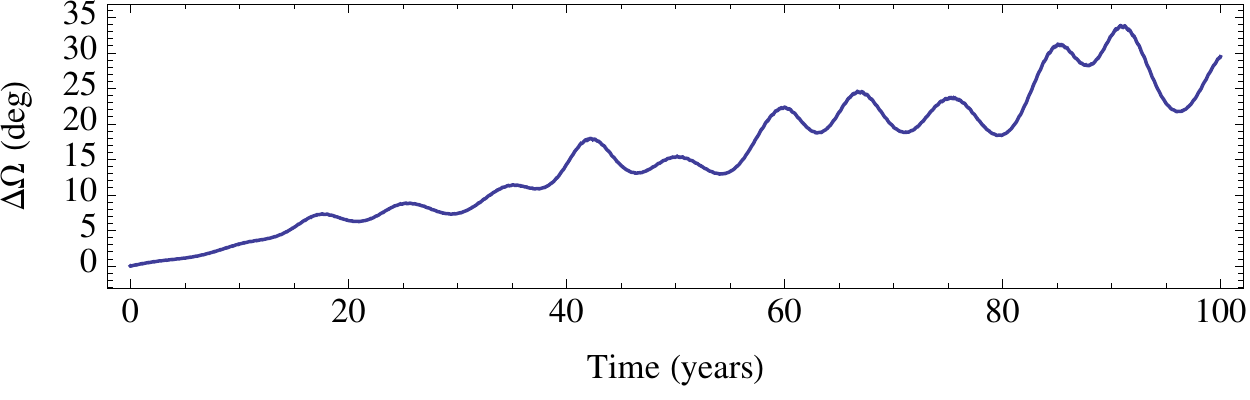} \\
\includegraphics[scale=0.6]{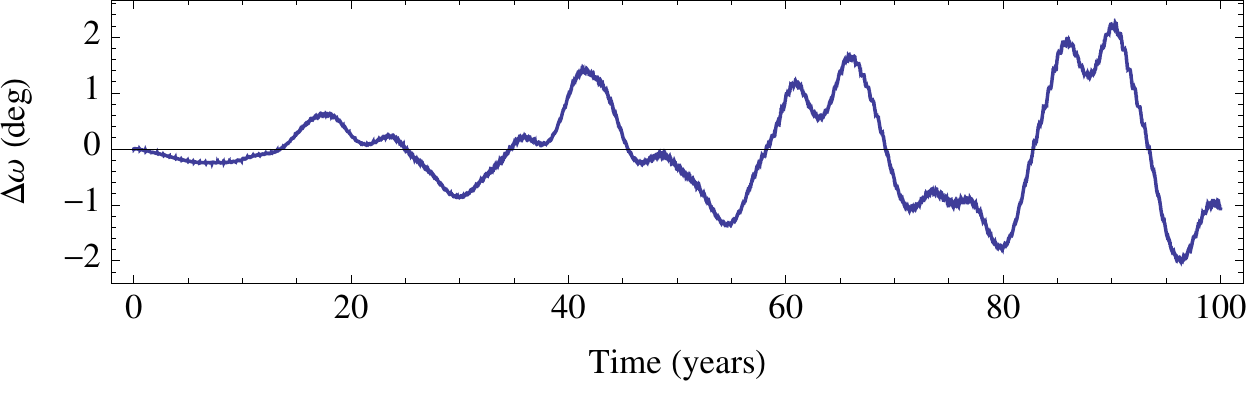} 
\caption{Errors between the numerical reference and the mean elements propagation: Molniya case. }
\label{f:Molniyaer}
\end{figure}

The differences between the mean elements provided by HEOSAT software and the osculating elements provided by the numerical reference are better appreciated in Fig.~\ref{f:Molniyaer}, where it is shown that periodic errors in the semi-major axes are of the order of 100 km. On the other hand, the more relevant discrepancies between the HEOSAT propagation and the numerical reference happen to the RAAN, in which case long-period errors of growing amplitude superimpose to a liner trend of $\sim0.3$ deg/year.
\par

\subsection{SimbolX orbit}

The second test presented is for a SymbolX type orbit, with initial conditions corresponding to the osculating elements 
\begin{equation}
\begin{array}{rcl}
a &=& 106247.136454\,\mathrm{km} \\
e &=& 0.75173 \\
I &=& 5.2789 \deg \\
\Omega &=& 49.351 \deg \\
\omega &=& -179.992 \deg \\
M &=& 0
\end{array}
\end{equation}
and the integration step size of the numerical reference is now 60 seconds. In this case the orbit apogee can reach half the earth-moon distance, and, therefore, the SymbolX orbit undergoes important third-body perturbations due to the moon's gravitational pull.
\par

\begin{figure}[htb]
\includegraphics[scale=0.6]{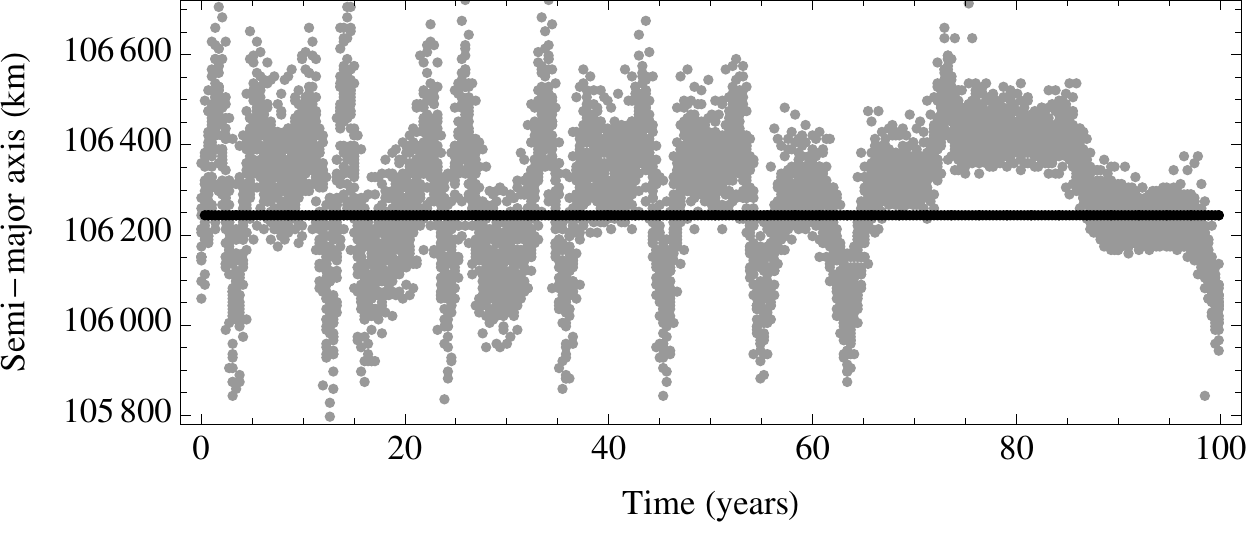} \\
\includegraphics[scale=0.6]{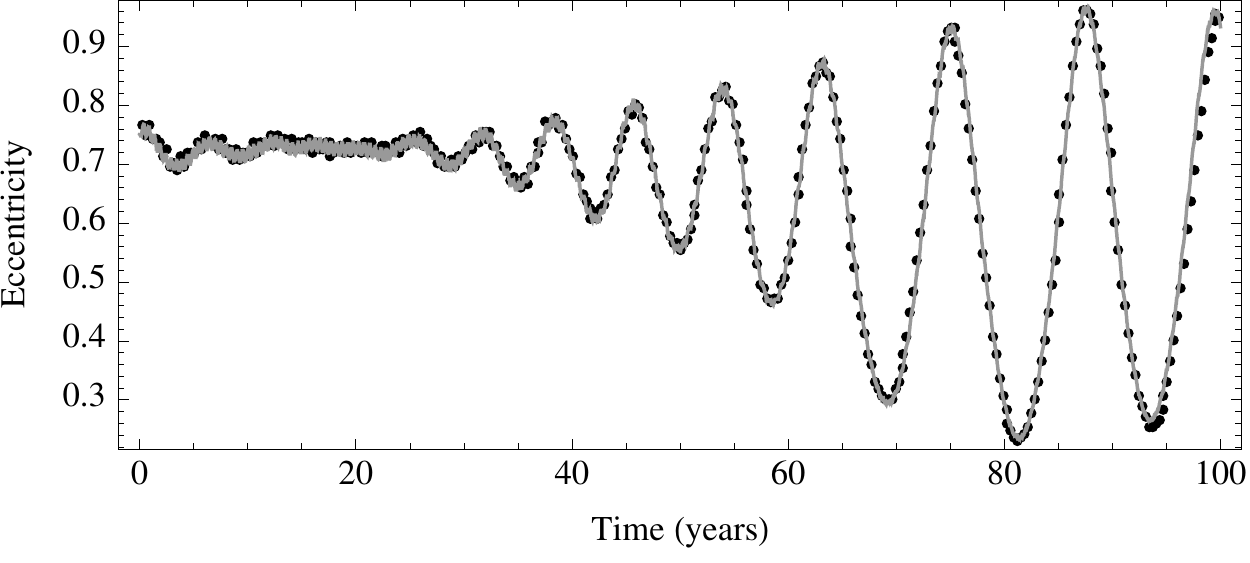} \\
\includegraphics[scale=0.6]{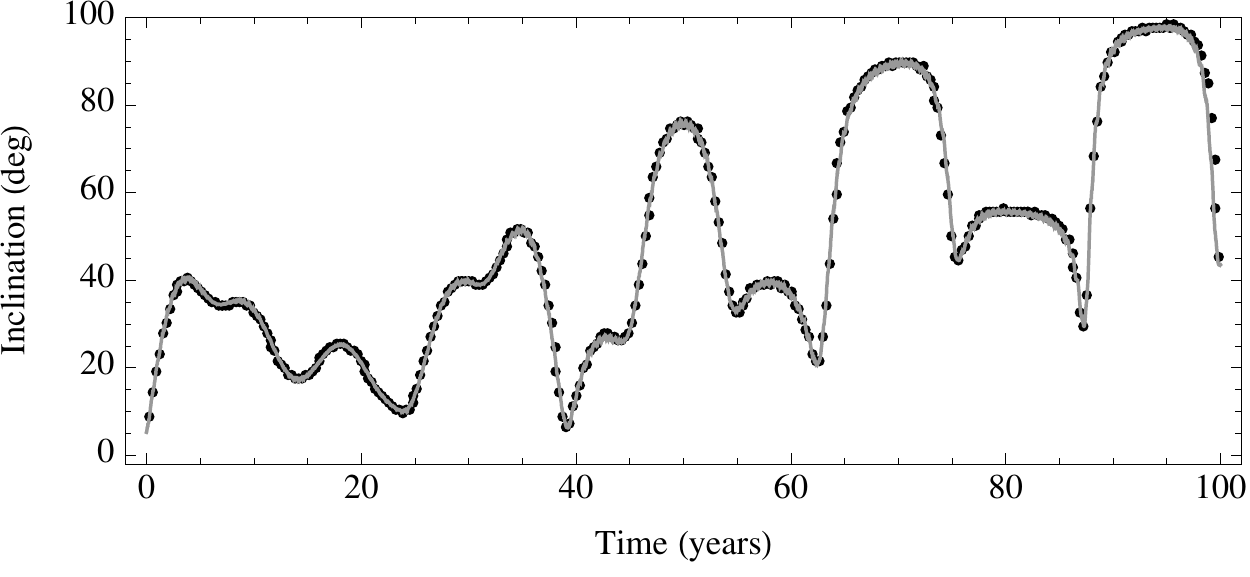} \\
\includegraphics[scale=0.6]{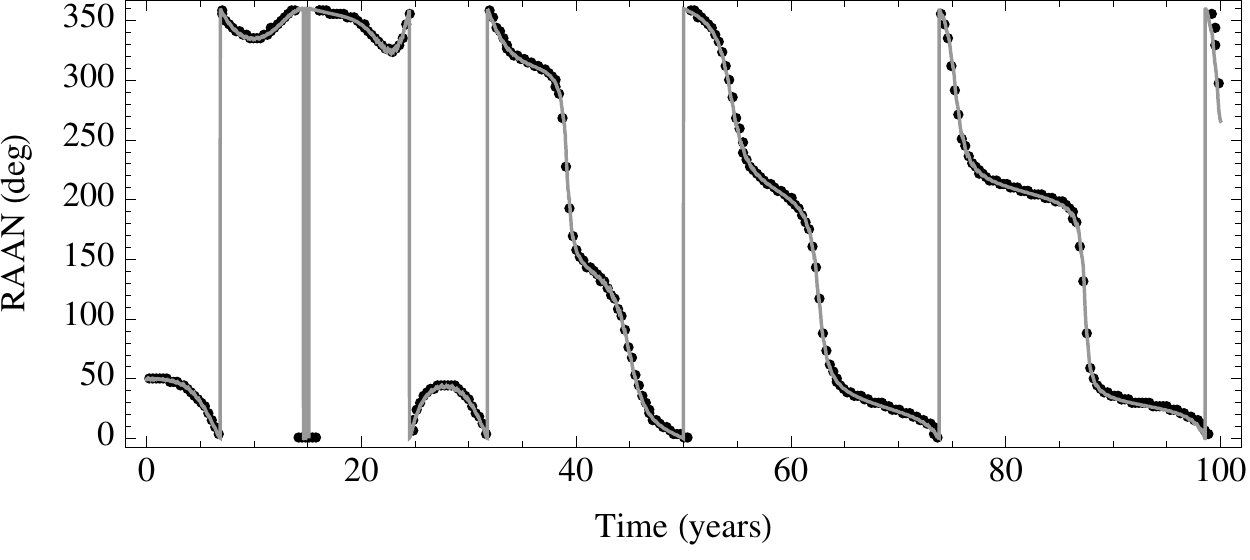} \\
\includegraphics[scale=0.6]{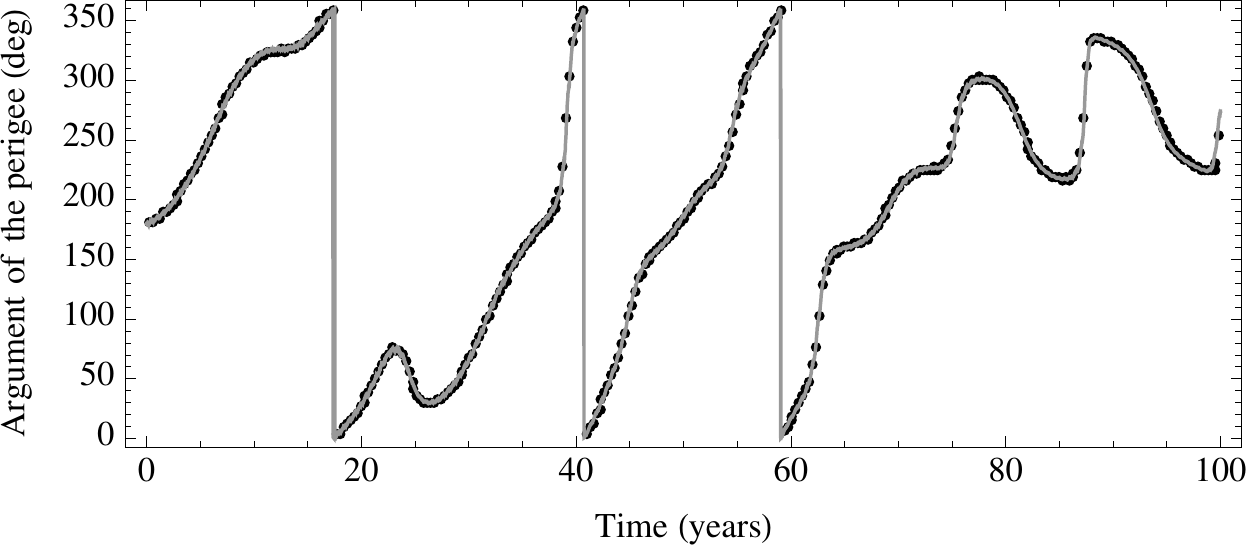} 
\caption{Time history of the orbit elements of the SymbolX. Dots: analytical propagation; gray line: numerical reference}
\label{f:SymbolXth}
\end{figure}

The time history of both the HEOSAT propagation and the numerically integrated reference are depicted in Fig.~\ref{f:SymbolXth}. As shown in the figure, the osculating semi-major axis experiences important variations whose amplitude can reach about 1000 km. These irregular variations are caused by the moon third-body perturbation, that make the SymbolX orbit to experience different resonances, which include a 1:7 resonance of the Laplace type, due to the orbital period of 4 days, as well as secular resonances of the Kozai type, cf.~\cite{Hautesserres2009}.
\par

\begin{figure}[htb]
\centering
\includegraphics[scale=0.6]{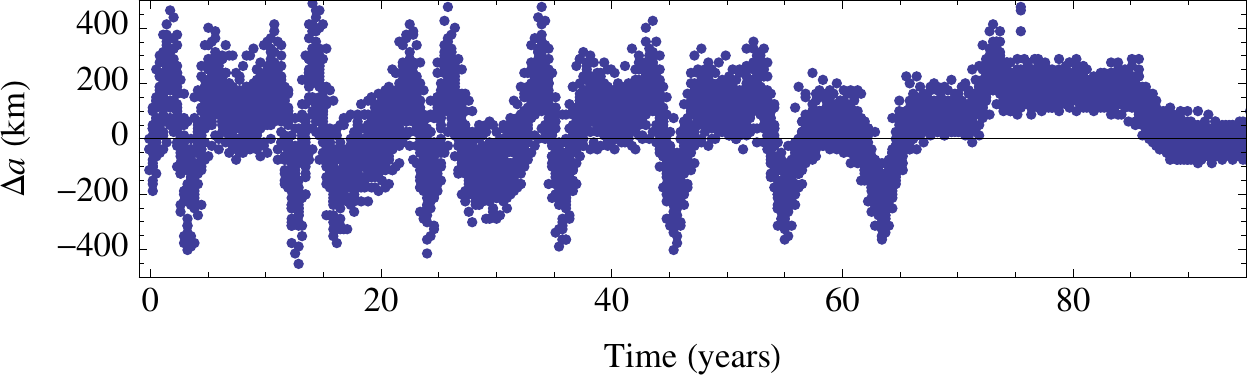} \\
\includegraphics[scale=0.6]{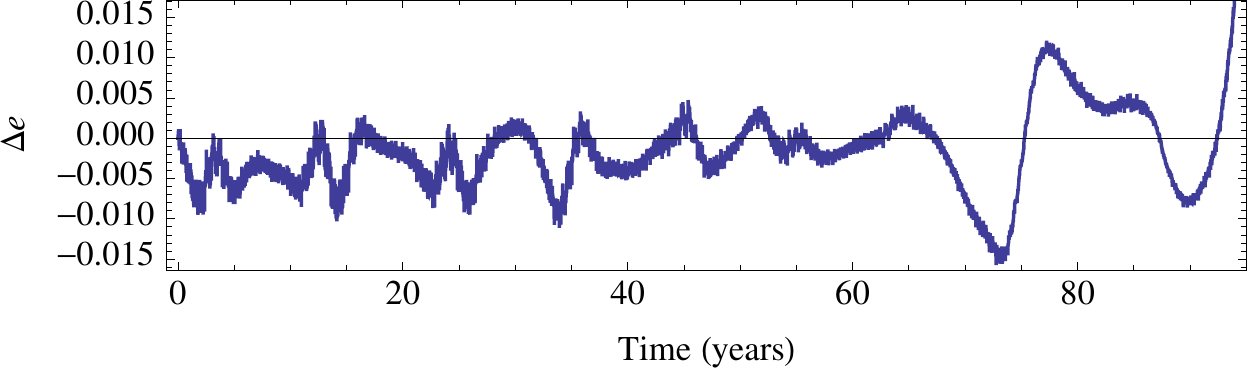} \\
\includegraphics[scale=0.6]{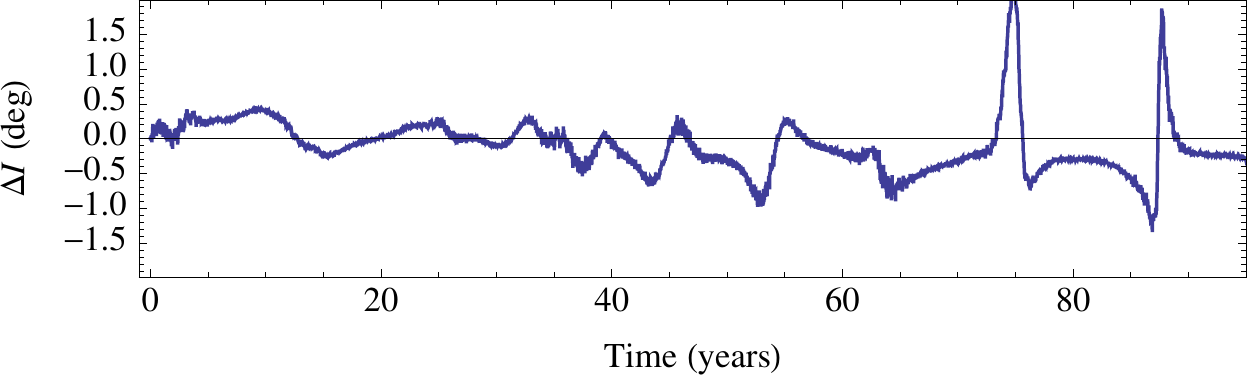} \\
\includegraphics[scale=0.6]{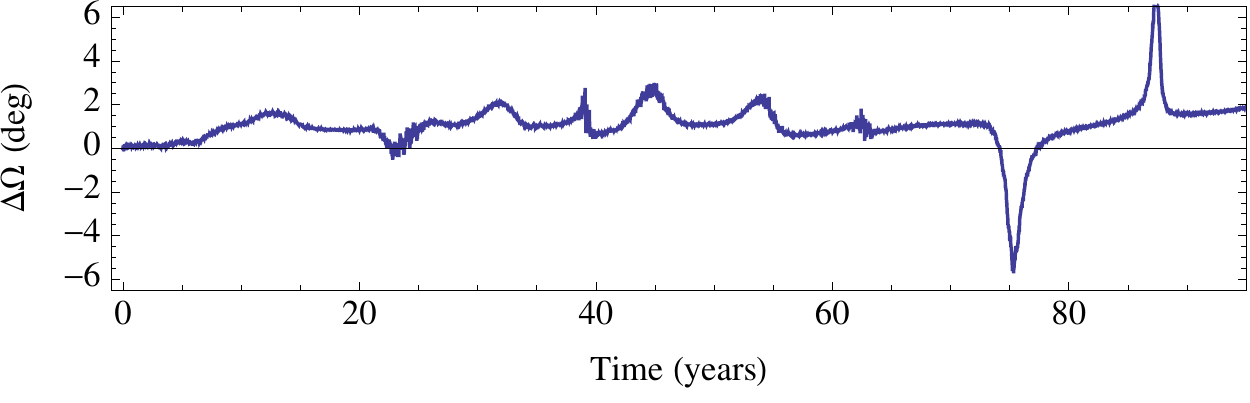} \\
\includegraphics[scale=0.6]{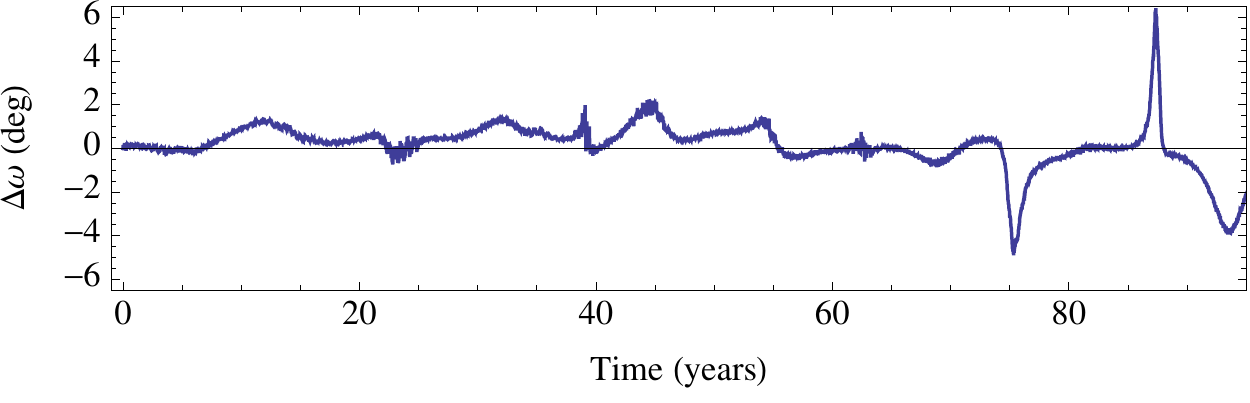} 
\caption{Errors between the numerical reference and the mean elements propagation: SymbolX case. }
\label{f:SymbolXer}
\end{figure}

Because of the irregularities in the time history of the osculating semi-major axis, we did not perform any adjustment in the computation of the HEOSAT mean semi-major axis, and the osculating initial elements are directly used as mean initial elements for launching the semi-analytical propagation. In spite of that, the time history of the mean orbital elements shows very good agreement with the osculating elements provided by the numerical reference. The detail on Fig.~\ref{f:SymbolXer} shows that this agreement extends for more than 70 years, and the discrepancies become important passed 95 years. Again, notable improvements are expected when the mean elements theory is completed with the analytic transformation from osculating to mean elements.

%


\section{Conclusions}

HEO propagation is a challenging problem due to the different perturbations that have an effect in highly elliptical orbits, the relative influence of which may notably vary along the orbit. However, modern tools and methods allow to approach the problem by means of analytical methods. Indeed, using perturbation theory we succeeded in the implementation of a fast and efficient semi-analytical propagator which is able to capture the main frequencies of the HEO motion over long time spans, even in extreme cases, as corroborated with the tests performed on the SymbolX orbit. In particular, we used the Lie transforms method, which is standard these days in the construction of perturbation theories. This method is specifically designed for automatic computation by machine, and is easily implemented with modern, commercial, general purpose software.
\par

Future evolutions of the semi-analytical theory should incorporate the transformation from osculating to mean elements, in this way enhancing the precision of the mean elements predictions based on it. Also, in spite of common HEO orbits are not affected by singularities, a reformulation in non-singular variables will make the orbit propagator software more versatile, widening its scope to the propagation of the majority of objects in a catalogue of earth satellite and debris orbits.

\begin{acknowledgements}
Partial support is acknowledged from the Ministry of Economic Affairs and Competitiveness of Spain, via Projects ESP2013-41634-P (M.L.) and ESP2014-57071-R (M.L. and J.F.S.).
\end{acknowledgements}

\appendix

\section{Some useful partial derivatives in elliptic motion} \label{ch:spd}

When dealing with automatic manipulation of literal expressions, it results practical to limit the symbolic algebra to the basic arithmetic operations, to wit, addition, subtraction, multiplication and division ---integer powers being a particular case of multiplication. However, square roots and trigonometric functions appear naturally in the formulation of the perturbation in canonical variables. To avoid dealing explicitly with square roots, it is wise to be equipped with a battery of partial derivatives that ease handling and simplifying symbolic expressions.
\par

Since our perturbation approach relies on the use of Delaunay and polar-nodal canonical variables, the partial derivatives of the classical Keplerian variables $(a,e,I,\Omega,\omega,M)$, standing for semi-major axis, eccentricity, inclination, right ascension of the ascending node, argument of the periapsis and mean anomaly, respectively, as well as the usual functions of the Keplerian variables
\begin{itemize}
\item the mean motion $n=\sqrt{\mu/a^3}$
\item the parameter (or \textit{semilatus rectum}) $p=a\,(1-e^2)$ 
\item the eccentricity function $\eta=\sqrt{1-e^2}$
\item the cosine of the inclination $c=\cos{I}$
\item the sine of the inclination $s=\sqrt{1-c^2}$
\item the eccentric anomaly $u$, given by $M=u-e\sin{u}$
\item the true anomaly $f$, given by $(1-e)\tan\frac{1}{2}f=\eta\tan\frac{1}{2}u$
\item the radial distance $r=a\,(1-e\cos{u})=p/(1+e\cos{f})$
\item the radial velocity $R=a\,n\,(a/r)\,e\sin{u}=a\,n\,(e/\eta)\sin{f}$
\item the argument of the latitude $\theta=f+\omega$
\item the projections of the eccentricity vector in the orbital frame: $k=e\cos{f}=-1+p/r$, and $q=e\sin{f}=R\,\eta/(n\,a)$
\end{itemize}
are provided both in the Delaunay chart $(\ell,g,h,L,G,H)$ and in the polar-nodal chart $(r,\theta,\nu,R,\Theta,N)$.
\par

We found convenient to express all the partial derivatives by means of the Keplerian functions:
\[
p,n,e,\eta,k,q,s,c
\]
from whose definition it is obtained
\begin{eqnarray*}
L &=& n\,p^2/\eta^4 \\
G &=& L\,\eta=n\,p^2/\eta^3=\Theta \\
H &=& G\,c=n\,p^2\,c/\eta^3=N \\
R &=& n\,p\,q/\eta^3 \\
r &=& p/(1+k)
\end{eqnarray*}
Recall also that, from the ellipse geometry, the following relations apply, cf.~Eq.~(\ref{ellipse}),
\[
\sin{f}=(a/r)\eta\sin{u}, \qquad \cos{f}=(a/r)(\cos{u}-e).
\]

Finally, it worths to mention that the use of logarithmic derivatives is helpful in finding the differentials that eased the computation of the partial derivatives. Note that only non-vanishing derivatives are presented.

\subsection{With respect to Delaunay variables}

\subsubsection{Orbital elements and related functions}
\begin{itemize}

\item Semi-major axis $a$:
\begin{equation}
\frac{\mathrm{d}a}{\mathrm{d}L}=2\frac{L}{\mu} = 2\frac{\eta ^2}{n\,p}
\end{equation}

\item Mean motion $n$:
\begin{equation}
\frac{\mathrm{d}n}{\mathrm{d}L} = -3\frac{\eta^4}{p^2},
\end{equation}

\item Parameter $p$:
\begin{equation}
\frac{\mathrm{d}p}{\mathrm{d}G}= 2\frac{G}{\mu} =2\frac{\eta^3}{n\,p}.
\end{equation}

\item Eccentricity function $\eta$:
\begin{eqnarray}
\frac{\mathrm{d}\eta}{\mathrm{d}L} &=& \eta\left(-\frac{1}{L}\right) = -\frac{\eta^5}{n\,p^2} \\
\frac{\mathrm{d}\eta}{\mathrm{d}G} &=& \eta\left(\frac{1}{G}\right) = \frac{\eta^4}{n\,p^2}
\end{eqnarray}

\item Eccentricity $e$:
\begin{eqnarray}
\frac{\mathrm{d}e}{\mathrm{d}L} &=& \frac{\eta^2}{e}\,\frac{1}{L} = \frac{\eta^6}{e\,n\,p^2} \\
\frac{\mathrm{d}e}{\mathrm{d}G} &=& -\frac{\eta^2}{e}\,\frac{1}{G} =-\frac{\eta^5}{e\,n\,p^2}
\end{eqnarray}

\item Cosine of inclination $c$:
\begin{eqnarray}
\frac{\mathrm{d}c}{\mathrm{d}G} &=& -\frac{c}{G}=-c\,\frac{\eta^3}{n\,p^2} \\
\frac{\mathrm{d}c}{\mathrm{d}H} &=& \frac{c}{H}=\frac{1}{G}=\frac{\eta^3}{n\,p^2}
\end{eqnarray}

\item Sine of inclination $s$:
\begin{eqnarray}
\frac{\mathrm{d}s}{\mathrm{d}G} &=& \frac{c^2}{s}\,\frac{\eta^3}{n\,p^2} \\
\frac{\mathrm{d}s}{\mathrm{d}H} &=& -\frac{c}{s}\,\frac{\eta^3}{n\,p^2}
\end{eqnarray}

\item Eccentric anomaly $u$:
\begin{eqnarray}
\frac{\mathrm{d}u}{\mathrm{d}\ell} &=& \frac{p}{r\,\eta^2}=\frac{1+k}{\eta^2} \\
\frac{\mathrm{d}u}{\mathrm{d}L} &=& 
\frac{R\,\eta^8}{e^2\,n^2\,p^3}
= \frac{q\,\eta^5}{e^2\,n\,p^2} \\
\frac{\mathrm{d}u}{\mathrm{d}G} &=&
-\frac{R\,\eta^7}{e^2\,n^2\,p^3}
=-\frac{q\,\eta^4}{e^2\,n\,p^2}
\end{eqnarray}

\end{itemize}

\subsubsection{Polar-nodal variables and related functions}

\begin{itemize}

\item Radial distance $r$:
\begin{eqnarray}
\frac{\mathrm{d}r}{\mathrm{d}\ell} &=&
\frac{R}{n}=\frac{p\,q}{\eta^3} \\
\frac{\mathrm{d}r}{\mathrm{d}L} &=&  
\frac{\eta^4}{e^2\,n}\left(\frac{2 e^2 r}{p^2}+\frac{1}{p}-\frac{1}{r}\right)
=\frac{\eta^4}{n\,p}\left(\frac{2}{1+k}-\frac{k}{e^2}\right)  \\
\frac{\mathrm{d}r}{\mathrm{d}G} &=&
\frac{\eta^3}{e^2\,n}\left(\frac{1}{r}-\frac{1}{p}\right)
=\frac{\eta^3\,k}{e^2\,n\,p}
\end{eqnarray}

\item Radial velocity $R$:
\begin{eqnarray}
\frac{\mathrm{d}R}{\mathrm{d}\ell} &=&
\frac{p\,n}{\eta^6}\frac{p^2}{r^2}\left(\frac{p}{r}-1\right)
=\frac{p\,n}{\eta^6}\,k\,(1+k)^2 \\
\frac{\mathrm{d}R}{\mathrm{d}L} &=& 
\eta^4\,\frac{R}{n\,r^2}\left(\frac{1}{e^2}-\frac{r^2}{p^2}\right)
=\eta\,\frac{q}{p}\left[\frac{(1+k)^2}{e^2}-1\right] \\
\frac{\mathrm{d}R}{\mathrm{d}G} &=&
-\frac{\eta^3}{e^2}\,\frac{R}{n\,r^2}
=-\frac{(1+k)^2}{e^2}\,\frac{q}{p} 
\end{eqnarray}

\item True anomaly $f$:
\begin{eqnarray}
\frac{\mathrm{d}f}{\mathrm{d}\ell} &=& \frac{\eta\,r}{a\,n\,(p-r)}\,\frac{\mathrm{d}R}{\mathrm{d}\ell}
=\frac{p^2}{r^2\,\eta^3}
=\frac{(1+k)^2}{\eta^3} \\
\frac{\mathrm{d}f}{\mathrm{d}L} &=& 
\frac{R\,\eta^7}{e^2\,n^2\,p^2}\left(\frac{1}{p}+\frac{1}{r}\right)=\frac{q\,\eta^4}{e^2\,n\,p^2}\,(2+k) \\
\frac{\mathrm{d}f}{\mathrm{d}G} &=& 
-\frac{R\,\eta^6}{e^2\,n^2\,p^2}\left(\frac{1}{p}+\frac{1}{r}\right)
=-\frac{q\,\eta^3}{e^2\,n\,p^2}\,(2+k)
\end{eqnarray}

\item Argument of the latitude $\theta$:
\begin{eqnarray}
\frac{\mathrm{d}\theta}{\mathrm{d}\ell} &=& \frac{\mathrm{d}f}{\mathrm{d}\ell}
\\
\frac{\mathrm{d}\theta}{\mathrm{d}g} &=& 1 \\
\frac{\mathrm{d}\theta}{\mathrm{d}L} &=& \frac{\mathrm{d}f}{\mathrm{d}L}
\\
\frac{\mathrm{d}\theta}{\mathrm{d}G} &=& \frac{\mathrm{d}f}{\mathrm{d}G}
\end{eqnarray}

\item Modulus of the angular momentum $\Theta$:
\begin{equation}
\frac{\mathrm{d}\Theta}{\mathrm{d}G} = 1
\end{equation}

\item Argument of the node $\nu$:
\begin{equation}
\frac{\mathrm{d}\nu}{\mathrm{d}h} = 1
\end{equation}

\item Polar component of the angular momentum $N$:
\begin{equation}
\frac{\mathrm{d}N}{\mathrm{d}H} = 1
\end{equation}

\item Eccentricity vector $k$:
\begin{eqnarray}
\frac{\mathrm{d}k}{\mathrm{d}\ell} &=& 
-\frac{p\,R}{r^2\,n}
=-\frac{q}{\eta^3}(1+k)^2 \\
\frac{\mathrm{d}k}{\mathrm{d}L} &=& 
\frac{\eta^4}{n\,r^2}\left(\frac{k}{e^2}-\frac{2}{k+1}\right) 
\\
\frac{\mathrm{d}k}{\mathrm{d}G} &=& 
-\frac{\eta^3}{n\,r^2}\left(\frac{k}{e^2}-\frac{2}{k+1}\right)
\end{eqnarray}

\item Eccentricity vector $q$:
\begin{eqnarray}
\frac{\mathrm{d}q}{\mathrm{d}\ell} &=&
\frac{1}{\eta^3}\frac{p^2}{r^2}\left(\frac{p}{r}-1\right)
=\frac{k}{\eta^3}\frac{p^2}{r^2}
=\frac{k}{\eta^3}\,(1+k)^2 \\
\frac{\mathrm{d}q}{\mathrm{d}L} &=&
\frac{q\,\eta^4}{n\,r^2}\left(\frac{1}{e^2}-\frac{r^2}{p^2}\right)
=\frac{q\,\eta^4}{n\,p^2}\left[\frac{(1+k)^2}{e^2}-1\right] \\
\frac{\mathrm{d}q}{\mathrm{d}G} &=&
-\frac{q\,\eta^3}{n\,r^2}\left(\frac{1}{e^2}-\frac{r^2}{p^2}\right)
=-\frac{q\,\eta^3}{n\,p^2}\left[\frac{(1+k)^2}{e^2}-1\right]\end{eqnarray}

\end{itemize}

\subsection{With respect to polar-nodal variables}

\subsubsection{Delaunay variables}

\begin{itemize}
\item Delaunay action $L$:
\begin{eqnarray}
\frac{\mathrm{d}L}{\mathrm{d}r} &=&
-\frac{\Theta}{p}\,\frac{k}{\eta^3}\,(1+k)^2
=-np\,\frac{k}{\eta^6}\,(1+k)^2 \\
\frac{\mathrm{d}L}{\mathrm{d}R} &=&
\frac{\Theta}{n}\,\frac{q}{p}
= p\frac{q}{\eta^3} \\
\frac{\mathrm{d}L}{\mathrm{d}\Theta} &=&
\frac{p^2}{\eta^3\,r^2}
= \frac{(1+k)^2}{\eta^3}
\end{eqnarray}

\item Modulus of the angular momentum $\Theta$:
\begin{equation}
\frac{\mathrm{d}G}{\mathrm{d}\Theta} = 1
\end{equation}

\item Polar component of the angular momentum $N$:
\begin{equation}
\frac{\mathrm{d}H}{\mathrm{d}N} = 1
\end{equation}

\item Mean anomaly $\ell$:
\begin{eqnarray}
\frac{\mathrm{d}\ell}{\mathrm{d}r} &=& \eta\,\frac{q}{r}\left(\frac{1+k}{e^2}-\frac{1}{1+k}\right) \\
\frac{\mathrm{d}\ell}{\mathrm{d}R} &=& \eta\,\frac{q}{R}\left(\frac{k}{e^2}-\frac{2}{1+k}\right) \\
\frac{\mathrm{d}\ell}{\mathrm{d}\Theta} &=& -\eta\,\frac{q}{\Theta}\,\frac{2+k}{e^2}
\end{eqnarray}

\item Argument of the perigee $g$:
\begin{eqnarray}
\frac{\mathrm{d}g}{\mathrm{d}r} &=& -\frac{q}{r}\,\frac{1+k}{e^2} \\
\frac{\mathrm{d}g}{\mathrm{d}\theta} &=& 1 \\
\frac{\mathrm{d}g}{\mathrm{d}R} &=& -\frac{q}{R}\,\frac{k}{e^2} \\
\frac{\mathrm{d}g}{\mathrm{d}\Theta} &=& \frac{q}{\Theta}\,\frac{2+k}{e^2}
\end{eqnarray}

\item Right ascension of the ascending node $h$:
\begin{eqnarray}
\frac{\mathrm{d}h}{\mathrm{d}\nu} &=& 1
\end{eqnarray}

\end{itemize}

\subsubsection{Orbital elements and related functions}

\begin{itemize}
\item Parameter $p$:
\begin{eqnarray}
\frac{\mathrm{d}p}{\mathrm{d}\Theta} &=& 2\frac{p}{\Theta}=2\frac{\eta^3}{n\,p}
\end{eqnarray}

\item Eccentricity vector $k$:
\begin{eqnarray}
\frac{\mathrm{d}k}{\mathrm{d}r} &=& -\frac{p}{r^2}=-\frac{(1+k)^2}{p} \\
\frac{\mathrm{d}k}{\mathrm{d}\Theta} &=&
\frac{2p}{r\,\Theta}=\frac{2q}{r\,R}
=\frac{2(1+k)\,\eta^3}{n\,p^2}
\end{eqnarray}

\item Eccentricity vector $q$:
\begin{eqnarray}
\frac{\mathrm{d}q}{\mathrm{d}R} &=& \frac{q}{R}=\frac{p}{\Theta}
=\frac{\eta^3}{n\,p} \\
\frac{\mathrm{d}q}{\mathrm{d}\Theta} &=& \frac{q}{\Theta}=\frac{q^2}{R\,p}
=\frac{q\,\eta^3}{n\,p^2}
\end{eqnarray}

\item Eccentricity $e$:
\begin{eqnarray}
\frac{\mathrm{d}e}{\mathrm{d}r} &=&
-\frac{k}{e}\,\frac{p}{r^2}
=-\frac{k}{e}\,\frac{(1+k)^2}{p} \\
\frac{\mathrm{d}e}{\mathrm{d}R} &=&
\frac{q^2}{e\,R}
=\frac{q\,\eta^3}{e\,n\,p} \\
\frac{\mathrm{d}e}{\mathrm{d}\Theta} &=&
\eta^3\frac{2k(1+k)+q^2}{e\,n\,p^2}
\end{eqnarray}

\item Eccentricity function $\eta$:
\begin{eqnarray}
\frac{\mathrm{d}\eta}{\mathrm{d}r} &=& \frac{k}{\eta}\,\frac{p}{r^2}
=\frac{k}{\eta}\,\frac{(1+k)^2}{p} \\
\frac{\mathrm{d}\eta}{\mathrm{d}R} &=& -\frac{q^2}{\eta\,R}
=-\frac{q\,\eta^2}{n\,p} \\
\frac{\mathrm{d}\eta}{\mathrm{d}\Theta} &=& =-\eta^2\frac{2k\,(1+k)+q^2}{n\,p^2}
\end{eqnarray}

\item Semi-major axis $a$:
\begin{eqnarray}
\frac{\mathrm{d}a}{\mathrm{d}r} &=&
-2\frac{k}{\eta^4}\,(1+k)^2 \\
\frac{\mathrm{d}a}{\mathrm{d}R} &=&
2\frac{p\,q^2}{R\,\eta^4}=2\frac{q}{n\,\eta} \\
\frac{\mathrm{d}a}{\mathrm{d}\Theta} &=&
2\frac{p^3}{\eta^4\,\Theta\,r^2} =2\frac{(1+k)^2}{n\,p\,\eta}
\end{eqnarray}

\item Mean motion $n$:
\begin{eqnarray}
\frac{\mathrm{d}n}{\mathrm{d}r} &=& \frac{3n\,k}{\eta^2}\,\frac{(1+k)^2}{p} \\
\frac{\mathrm{d}n}{\mathrm{d}R} &=& -\frac{3n\,q^2}{R\,\eta^2}=-3\frac{q\,\eta}{p} \\
\frac{\mathrm{d}n}{\mathrm{d}\Theta} &=& -\frac{3n\,p^2}{\eta^2\,\Theta\,r^2} =-\frac{3\eta}{r^2}
\end{eqnarray}

\item True anomaly $f$:
\begin{eqnarray}
\frac{\mathrm{d}f}{\mathrm{d}r} &=& \frac{q}{r}\,\frac{1+k}{e^2} \\
\frac{\mathrm{d}f}{\mathrm{d}R} &=& \frac{q}{R}\,\frac{k}{e^2} \\
\frac{\mathrm{d}f}{\mathrm{d}\Theta} &=& -\frac{q}{\Theta}\,\frac{2+k}{e^2}
\end{eqnarray}

\item Eccentric anomaly $u$:
\begin{eqnarray}
\frac{\mathrm{d}u}{\mathrm{d}r}
&=&
\frac{\eta}{pq}\left[1+\frac{e^2-k}{e^2}\frac{k}{\eta^2}\,(1+k)\right](1+k) \\
\frac{\mathrm{d}u}{\mathrm{d}R} &=& \frac{a\,n}{R}\left(-\frac{r-a}{r}\,\frac{\mathrm{d}e}{e\mathrm{d}R}-\frac{\mathrm{d}a}{a\mathrm{d}R}\right)
\\
\frac{\mathrm{d}u}{\mathrm{d}\Theta} &=& \frac{a\,n}{R}\left(-\frac{r-a}{r}\,\frac{\mathrm{d}e}{e\mathrm{d}\Theta}-\frac{\mathrm{d}a}{a\mathrm{d}\Theta}\right)
\end{eqnarray}

\item Equation of the center $\phi$:
\begin{eqnarray}
\frac{\mathrm{d}\phi}{\mathrm{d}r} &=& \frac{q}{r}\left(\frac{1+k}{1+\eta}+\frac{\eta}{1+k}\right) \\
\frac{\mathrm{d}\phi}{\mathrm{d}R} &=& \frac{q}{R}\left(\frac{k}{1+\eta}+\frac{2\eta}{1+k}\right) \\
\frac{\mathrm{d}\phi}{\mathrm{d}\Theta} &=& -\frac{q}{\Theta}\,\frac{2+k}{1+\eta}
\end{eqnarray}

\item Cosine of inclination $c$:
\begin{eqnarray}
\frac{\mathrm{d}c}{\mathrm{d}\Theta} &=& -\frac{c}{\Theta} =-\frac{c\,q}{R\,p}
= -\frac{c\,\eta^3}{n\,p^2} \\
\frac{\mathrm{d}c}{\mathrm{d}N} &=& \frac{1}{\Theta} =\frac{q}{R\,p}
=\frac{\eta^3}{n\,p^2}
\end{eqnarray}

\item Sine of inclination $s$:
\begin{eqnarray}
\frac{\mathrm{d}s}{\mathrm{d}\Theta} &=&
 \frac{c^2\,\eta^3}{s\,n\,p^2} \\
\frac{\mathrm{d}s}{\mathrm{d}N} &=&
-\frac{c\,\eta^3}{s\,n\,p^2}
\end{eqnarray}

\end{itemize}


\section{Tables of coefficients} \label{ap:tables}

The coefficients of the trigonometric series used by HEOSAT are provided in following tables

\begin{table*}[htbp]
\caption{Eccentricity polynomials $Q_{m,k+m\bmod2}$ in Eq.~(\protect\ref{Ji}); $Q_{m,m-2}=1$, $Q_{m,m-4}=2m-6+3e^2$ and $Q_{6,0}=8 + 40 e^2 + 15 e^4$. \label{t:EPzonal} }
\begin{tabular}{@{}llllll@{}}
$k$ & \multicolumn{1}{c}{$m=7$} & \multicolumn{1}{c}{$m=8$} 
    & \multicolumn{1}{c}{$m=9$} & \multicolumn{1}{c}{$m=10$} 
\\ [0.33ex]
\hline
$0$ & $3 (8 + 20 e^2 + 5 e^4)$ & $3 (16 + 168 e^2 + 210 e^4 + 35 e^6)$
    & $3 (64 + 336 e^2 + 280 e^4 + 35 e^6)$ & $3 (128 + 2304 e^2 + 6048 e^4 + 3360 e^6 + 315 e^8)$ \vphantom{$\frac{M^9}{M^9}$} \\ [0.4ex]
$2$ & & $5 (16 + 20 e^2 + 3 e^4)$ & $48 + 80 e^2 + 15 e^4$ & $15 (32 + 112 e^2 + 70 e^4 + 7 e^6)$ \\ [0.4ex]
$4$ & & & & $15 (8 + 8 e^2 + e^4)$ \\ [0.4ex]
\hline
\end{tabular}
\end{table*}
\begin{table*}[htb]
\caption{Even inclination polynomials $B_{2m,2k}$ in Eq.~(\protect\ref{Ji}). 
\label{t:IPzeven}}
\begin{tabular}{@{}lllll@{}}
$k$ & \multicolumn{1}{c}{$m=1$} & \multicolumn{1}{c}{$m=2$} & \multicolumn{1}{c}{$m=3$} & \multicolumn{1}{c}{$m=4$} \\
\hline
$0$ & $\frac{1}{4}\left(3 c^2-1\right)$ & $-\frac{3}{128}\left(35 c^4-30 c^2+3\right)$
    & $\frac{5}{2048}\left(231 c^6-315 c^4+105 c^2-5\right)$
    & $-\frac{35}{786432}\left(6435 c^8-12012 c^6+6930 c^4-1260c^2+35\right)$ \vphantom{$\frac{M^9}{M^9}$} \\ [0.4ex]
$1$ & & $-\frac{15}{64} \left(7c^2-1\right)$ & $\frac{175}{2048}\left(33 c^4-18 c^2+1\right)$
    & $-\frac{2205}{131072}\left(143 c^6-143 c^4+33c^2-1\right)$ \\ [0.4ex]
$2$ & & & $\frac{315}{4096}\left(11c^2-1\right)$ & $-\frac{4851}{131072}\left(65 c^4-26c^2+1\right)$ \\ [0.4ex]
$3$ & & & & $-\frac{3003}{131072}\left(15 c^2-1\right)$ \\ [0.4ex]
\hline
 & \multicolumn{3}{c}{$m=5$} \\
\hline
$0$ &  \multicolumn{3}{l}{$\frac{21}{8388608}\left(46189 c^{10}-109395 c^8+90090 c^6-30030 c^4+3465c^2-63\right)$} \vphantom{$\frac{M^9}{M^9}$} \\ [0.4ex]
$1$ &  \multicolumn{3}{l}{$\frac{693}{2097152}\left(4199 c^8-6188 c^6+2730c^4-364 c^2+7\right)$} \\ [0.4ex]
$2$ &  \multicolumn{3}{l}{$\frac{9009}{1048576}\left(323 c^6-255 c^4+45c^2-1\right)$} \\ [0.4ex]
$3$ &  \multicolumn{3}{l}{$\frac{19305}{4194304}\left(323 c^4-102c^2+3\right)$} \\ [0.4ex]
$4$ &  \multicolumn{3}{l}{$\frac{109395}{16777216}\left(19c^2-1\right)$} \\ [0.4ex]
\hline
\end{tabular}
\end{table*}
\begin{table*}[htb]
\caption{Odd inclination polynomials $B_{2m+1,2k+1}$ in Eq.~(\protect\ref{Ji}). \label{t:IPzodd} }
\begin{tabular}{@{}lllll@{}}
$k$ & \multicolumn{1}{c}{$m=1$} & \multicolumn{1}{c}{$m=2$} & \multicolumn{1}{c}{$m=3$} & \multicolumn{1}{c}{$m=4$} \\
\hline
$0$ & $-\frac{3}{8} \left(5 c^2-1\right)$ & $\frac{15}{128} \left(21 c^4-14 c^2+1\right)$
    & $-\frac{35}{8192}\left(429 c^6-495 c^4+135 c^2-5\right)\!\!$
    & $\frac{105}{262144}\left(2431 c^8-4004 c^6+2002 c^4-308c^2+7\right)$ \vphantom{$\frac{M^9}{M^9}$} \\ [0.4ex]
$1$ & & $\frac{35}{256} \left(9c^2-1\right)$ & $-\frac{315}{16384}\left(143 c^4-66 c^2+3\right)$ 
    & $\frac{1617}{131072}\left(221 c^6-195 c^4+39c^2-1\right)$ \\ [0.4ex]
$2$ & & & $-\frac{693}{16384}\left(13c^2-1\right)$ & $\frac{3003}{131072}\left(85 c^4-30c^2+1\right)$ \\ [0.4ex]
$3$ & & & & $\frac{6435}{524288}\left(17 c^2-1\right)$ \\ [0.4ex]
\hline
\end{tabular}
\end{table*}
\begin{table*}[htbp]
\caption{Eccentricity polynomials $A_{m,j}$ in Eq.~(\protect\ref{gammam}). \label{t:EP} }
\begin{tabular}{@{}lllll@{}}
$m$ & \multicolumn{1}{l}{$j=0$} & \multicolumn{1}{l}{$1$} & \multicolumn{1}{l}{$2$} & \multicolumn{1}{c}{$3$}  \\[0.33ex]
\hline
$2$ & $3 (2 + 3 e^2)$ & $-15 e^2$ \vphantom{$\frac{M^9}{M^9}$} \\ [0.4ex]
$3$ & $e (4 + 3 e^2)$ & $e^3$ \\ [0.4ex]
$4$ & $(8 + 40 e^2 + 15 e^4)$ & $e^2 (2 + e^2)$ & $e^4$ \\ [0.4ex]
$5$ & $e (8 + 20 e^2 + 5 e^4)$ & $e^3 (8 + 3 e^2) $ & $e^5$ \\ [0.4ex]
$6$ & $16 + 168 e^2 + 210 e^4 + 35 e^6$ & $48e^2+80e^4+15e^6$ & $10e^4+3e^6$ & $e^6$ \\ [0.4ex]
\hline
\end{tabular}
\end{table*}
\begin{table*}[htbp]
\caption{Inclination polynomials $P_{m,j,l}$ in Eq.~(\protect\ref{gammam}) ($\chi=c\pm1$). \label{t:P3B2y3} }
\begin{tabular}{@{}lllcllclll@{}}
 & \multicolumn{2}{c}{$P_{2,j,l}$} && \multicolumn{2}{c}{$P_{3,j,l}$} && \multicolumn{3}{c}{$P_{4,j,l}$}   \\[0.33ex]
\hline
$\hphantom{\pm}l$ & \multicolumn{1}{c}{$j=0$} & \multicolumn{1}{c}{$j=1$} && \multicolumn{1}{c}{$j=0$} & \multicolumn{1}{c}{$j=1$}
 && \multicolumn{1}{c}{$j=0$} & \multicolumn{1}{c}{$j=1$} & \multicolumn{1}{c}{$j=2$} \\[0.33ex]
\hline
$\hphantom{+}0$ & $\frac{1}{48}(3 c^2-1)$ & $-\frac{1}{16}s^2$ & \hspace{0.2cm}
                & $-\frac{15}{128} (5 c^2-1) s$ & $-\frac{175}{128}s^3$ &  \hspace{0.2cm}
                & $-\frac{3}{4096}(35 c^4-30 c^2+3)$ & $-\frac{105}{1024}(7 c^2-1) s^2$ & $-\frac{2205}{4096} s^4$
                  \vphantom{$\frac{M^M}{M}$} \\ [0.4ex]
$\pm1$ & $\frac{1}{8}c s$ & $\frac{1}{8} \chi s$ &
       & $-\frac{15}{512} \chi(15 c^2\mp10 c-1)$ & $-\frac{525}{512} \chi s^2$ & 
       & $\frac{15}{1024}c(3-7 c^2) s$ & $\frac{105}{512} \chi(14 c^2\mp7 c-1) s$ & $\frac{2205}{1024}\chi s^3$ \\ [0.4ex]
$\pm2$ & $-\frac{1}{32}s^2$ & $\frac{1}{32} \chi^2$ &
       & $\frac{75}{256} \chi (3 c\mp1) s$ & $-\frac{525}{256} \chi^2 s$ &
       & $\frac{15}{1024}(7 c^2-1) s^2$ & $\frac{105}{256} \chi^2(7 c^2\mp7 c+1)$ & $\frac{2205}{1024}\chi^2 s^2$ \\ [0.4ex]
$\pm3$ & & & & $\frac{75}{512} \chi s^2$ & $\frac{175}{512} \chi^3$ &
       & $\frac{105}{1024}c s^3$ & $\frac{735}{512} \chi^2 (2 c\mp1) s$ & $-\frac{2205}{1024}\chi^3 s$ \\ [0.4ex]
$\pm4$ & & & & & & & $-\frac{105}{2048}s^4$ & $-\frac{735}{512} \chi^2 s^2$ & $-\frac{2205}{2048} \chi^4$ \\ [0.4ex]
\hline
\end{tabular}
\end{table*}

\begin{table*}[htbp]
\caption{Inclination polynomials $P_{5,j,l}$ in Eq.~(\protect\ref{gammam}) ($\chi=c\pm1$). \label{t:P3B5}}
\begin{tabular}{@{}llll@{}}
$\hphantom{\pm}l$ & \multicolumn{1}{c}{$j=0$} & \multicolumn{1}{c}{$j=1$} & \multicolumn{1}{c}{$j=2$}   \\[0.33ex]
\hline
$\hphantom{+}0$ & $\frac{105}{8192} \left(21 c^4-14 c^2+1\right)s$
       & $\frac{735}{16384}(9 c^2-1)s^3$ & $\frac{14553}{16384} s^5$ \vphantom{$\frac{M^M}{M}$} \\ [0.4ex]
$\pm1$ & $\frac{105}{16384}\chi(105 c^4\mp84c^3-42 c^2\pm28c+1)$
       & $\frac{2205}{32768}\chi(15 c^2\mp6 c-1)s^2$ & $\frac{72765}{32768} \chi s^4$ \\ [0.4ex]
$\pm2$ & $-\frac{735}{4096}\chi(15 c^3\mp9 c^2-3 c\pm1)s$
       & $\frac{2205}{8192}\chi^2(15 c^2\mp12 c+1)s$ & $\frac{72765}{8192} \chi^2 s^3$ \\ [0.4ex]
$\pm3$ & $\frac{735}{32768}\chi(15 c^2\mp6 c-1)s^2$
       & $\frac{735}{65536}\chi^3(3 c\mp1) (15c\mp13)$ & $\frac{72765}{65536} \chi^3 s^2$ \\ [0.4ex]
$\pm4$ & $-\frac{2205}{16384}\chi(5c\mp1) s^3$
       & $-\frac{6615}{32768}\chi^3(5c\mp3) s$ & $\frac{72765}{32768} \chi^4s$ \\ [0.4ex]
$\pm5$ & $\frac{2205}{32768}\chi s^4$
       & $\frac{6615}{65536}\chi^3s^2$ & $\frac{14553}{65536} \chi^5$ \\ [0.4ex]
\hline
\end{tabular}
\end{table*}

\begin{table*}[htbp]
\caption{Inclination polynomials $P_{6,j,l}$ in Eq.~(\protect\ref{gammam}) ($\chi=c\pm1$). \label{t:P602}}
\begin{tabular}{@{}lllll@{}}
$\hphantom{\pm}l$ & \multicolumn{1}{c}{$j=0$} & \multicolumn{1}{c}{$j=1$} & \multicolumn{1}{c}{$j=2$} & \multicolumn{1}{c}{$j=3$}   \\
\hline
$\hphantom{+}0$ & $-\frac{5}{65536}(231 c^6-315 c^4+105c^2-5)$ & $-\frac{315}{131072}(33 c^4-18 c^2+1)s^2$
                & $-\frac{2079}{65536}(11c^2-1)s^4$ & $-\frac{99099}{131072}s^6$ \vphantom{$\frac{M^M}{M}$} \\ [0.4ex]
$\pm1$ & $-\frac{105}{32768}c(33 c^4-30 c^2+5)s$ & $\hphantom{+}\frac{315}{65536}\chi(99 c^4\mp66 c^3-36 c^2\pm18c+1) s$
       & $\hphantom{+}\frac{2079}{32768}\chi(33 c^2\mp11 c-2) s^3$ & $\hphantom{+}\frac{297297}{65536}\chi s^5$ \\ [0.4ex]
$\pm2$ & $-\frac{525}{262144}(33 c^4-18 c^2+1)s^2$ & $-\frac{315}{524288}\chi^2(495 c^4\mp660 c^3+90 c^2\pm108c-17)$
       & $-\frac{10395}{262144}\chi^2(33 c^2\mp22 c+1) s^2$ & $-\frac{1486485}{524288}\chi^2s^4$ \\ [0.4ex]
$\pm3$ & $-\frac{525}{65536}c(11 c^2-3)s^3$ & $-\frac{945}{131072}\chi^2(55 c^3\mp55 c^2+5 c\pm3)s$
       & $\hphantom{+}\frac{10395}{65536} \chi ^3(11 c^2\mp11 c+2) s$ & $\hphantom{+}\frac{495495}{131072}\chi^3s^3$ \\ [0.4ex]
$\pm4$ & $\hphantom{+}\frac{315}{32768}(1-11c^2)s^4$ & $-\frac{945}{65536}\chi^2(33 c^2\mp22 c+1)s^2$
       & $-\frac{2079}{32768} \chi ^4(33 c^2\mp44 c+13)$ & $-\frac{297297}{65536}\chi^4s^2$ \\ [0.4ex]
$\pm5$ & $-\frac{3465}{65536}cs^5$ & $\hphantom{+}\frac{10395}{131072}(1\mp3c)\chi^2s^3$
       & $-\frac{22869}{65536} \chi ^4 (3 c\mp2)s$ & $\hphantom{+}\frac{297297}{131072}\chi ^5s$ \\ [0.4ex]
$\pm6$ & $\hphantom{+}\frac{1155}{262144}s^6$ & $\hphantom{+}\frac{10395}{524288}\chi^2 s^4$
       & $\hphantom{+}\frac{22869}{262144}\chi^4s^2$ & $\hphantom{+}\frac{99099}{524288}\chi ^6$ \\ [0.4ex]
\hline
\end{tabular}
\end{table*}

\begin{table*}[htbp]
\caption{Third-body direction polynomials in Eq.~(\protect\ref{gammam}); $u\equiv{u}^\star$, $v\equiv{v}^\star$, $w\equiv{w}^\star$. \label{t:3BP}}
\begin{tabular}{@{}lrlll@{}}
$m$ & $l$ & \multicolumn{1}{l}{$S_{m,l}$} & \multicolumn{1}{l}{$T_{m,l}$}  \\
\hline
$2$ & $0$ & $-1+3w^2$ & $0$ \\
 & $\pm1$ & $-v\,w$ & $\pm uw$ \\
 & $\pm2$ & $u^2-v^2$ & $\pm2u\,v$ \\
\hline
$3$ & $0$ & $0$ & $w (5 w^2-3)$ \\
 & $\pm1$ & $\pm u (5 w^2-1)$ & $v (5 w^2-1)$ \\
 & $\pm2$ & $\pm2uvw$ & $w (v^2-u^2)$ \\
 & $\pm3$ & $\pm u(u^2-3v^2)$ & $-v (v^2-3 u^2)$ \\
\hline
$4$ & $0$ & $3 - 30 w^2 + 35 w^4$ & $0$ \\
 & $\pm1$ & $vw(3-7w^2)$ & $\pm u\,w\,(-3+ 7 w^2)$ \\
 & $\pm2$ & $\frac{1}{2}(u^2-v^2)(-1+7w^2)$ & $\pm uv(-1+7w^2)$ \\
 & $\pm3$ & $v(-3u^2+v^2)w$ & $\pm u(u^2-3v^2)w$ \\
 & $\pm4$ & $\frac{1}{4}(u^4 - 6 u^2 v^2 + v^4)$ & $\pm uv(u^2-v^2)$ \\
\hline
$5$ & $0$ & $0$ & $w\,(15 - 70 w^2 + 63 w^4)$ \\
 & $\pm1$ & $\pm u(1-14w^2+21w^4)$ & $v(1-14w^2+21w^4)$ \\
 & $\pm2$ & $\pm 2uvw(-1+3w^2)$ & $(u^2-v^2)w(1-3w^2)$ \\
 & $\pm3$ & $\pm u(u^2-3v^2)(1-9w^2)$ & $v(-3u^2+v^2)(-1+9w^2)$ \\
 & $\pm4$ & $\pm 4uv(-u^2+v^2)w$ & $(u^4-6u^2v^2+v^4)w$  \\
 & $\pm5$ & $\pm u(u^4-10u^2v^2+5v^4)$ & $v(5u^4-10u^2v^2+v^4)$ \\
\hline
$6$ & $0$ & $ 5 - 105 w^2 + 315 w^4 - 231 w^6$ & $0$  \\
 & $\pm1$ & $v (5 - 30 w^2 + 33 w^4) w$ & $\mp u (5 - 30 w^2 + 33 w^4) w$ \\
 & $\pm2$ & $(u^2 - v^2) (1 - 18 w^2 + 33 w^4)$ & $\pm2 u v (1 - 18 w^2 + 33 w^4)$ \\
 & $\pm3$ & $v (3 u^2 - v^2) (3-11w^2) w$ & $\pm u (u^2 - 3 v^2) (-3 + 11 w^2) w$ \\
 & $\pm4$ & $\frac{1}{4}(1-11w^2) (u^4 - 6 u^2 v^2 + v^4)$ & $\pm(1-11w^2) u v (u^2 - v^2)$ \\
 & $\pm5$ & $v (5 u^4 - 10 u^2 v^2 + v^4) w$ & $\mp u (u^4 - 10 u^2 v^2 + 5 v^4) w$ \\
 & $\pm6$ & $(v^2-u^2)(u^4-14 u^2 v^2+v^4)$ & $\mp2(3u^5-10u^3v^2+3uv^4)v$ \\
\hline
\end{tabular}
\end{table*}

\end{document}